# Improving student understanding of quantum measurement in infinite-dimensional Hilbert space using a research-based multiple-choice question sequence

Yangqiuting Li [1] and Chandralekha Singh [2]

[1]*Department of Physics, Oregon State University, Corvallis, Oregon 97331, USA*
[2]*Department of Physics and Astronomy University of Pittsburgh, Pittsburgh PA 15260, USA*

## Abstract

Research-based multiple-choice questions implemented in class with peer instruction have been shown to be an effective tool for improving students' engagement and learning outcomes. Moreover, multiple- choice questions that are carefully sequenced to build on each other can be particularly helpful for students to develop a systematic understanding of concepts pertaining to a theme. Here, we discuss the development, validation, and implementation of a multiple-choice question sequence (MQS) on the topic of quantum measurement in the context of wave functions in the infinite-dimensional Hilbert space. This MQS was developed using students' common difficulties with quantum measurements as a guide and was implemented in a junior-/senior-level quantum mechanics course at a large research university in the U.S. We compare student performance on assessment tasks focusing on quantum measurement before and after the implementation of the MQS and discuss how different difficulties were reduced and how to further improve students' conceptual understanding of quantum measurement in infinite-dimensional Hilbert space.

## I. Introduction

Prior studies have shown that learning quantum mechanics is challenging for students at all levels, including advanced undergraduate and graduate students. Some studies have revealed that students often struggle with the abstract nature of quantum concepts and the visualization of nonintuitive phenomena [1–4]. Additionally, certain topics such as quantum measurements, time evolution, and the probabilistic framework of quantum mechanics are particularly challenging for students [5–8]. Furthermore, it has been found that students often overgeneralize concepts and have difficulties with the complex mathematical formalism of quantum mechanics, including solving partial differential equations and understanding the mathematical representation of wave functions in potential wells [5,9–11]. Many prior studies have focused on developing pedagogical strategies and curricular reforms aimed at fostering a deeper and more intuitive understanding of quantum mechanics [7,8,11–25].

The Physics Education Research group at the University of Pittsburgh has been involved in investigating student difficulties in learning quantum mechanics and using them as a guide to develop and validate curricula and pedagogies to reduce these difficulties. Our group's research spans a range of topics, for example, we investigated student understanding of quantum notation and mathematical formalism [26–32], quantum operators and observables [33–36], quantum information and interference [37–41], and perturbation theory [42–45]. Among the different topics undergraduates learn in quantum mechanics, quantum measurement is one of the key concepts.

The outcome of a quantum measurement is in general probabilistic and reflects the probabilistic nature of quantum mechanics. Quantum measurement is also fundamental to quantum computation since measurement is necessary to extract the outcome at the end of the computation. In prior studies, quantum measurement has been found to be one of the particularly difficult topics for students [7,36,46–50].

One reason why learning quantum measurement is challenging is the fact that concepts related to quantum measurement are very different from those in classical mechanics, with which students are likely to be familiar [32]. In classical mechanics, physical observables such as position and energy have well-defined values even before the measurement. However, in a generic quantum state, physical observables such as position and energy are usually not well defined, and there are many possible outcomes that can be measured for an observable [51]. Also, for a quantum system, each observable corresponds to a Hermitian operator, which has a complete set of orthonormal eigenstates and a corresponding set of real eigenvalues. Any quantum state of this system can be expanded as a linear superposition of the complete set of eigenstates of the operator corresponding to any observable. When an observable is measured, the quantum state instantaneously collapses into an eigenstate of the operator corresponding to the measured observable and we obtain a corresponding eigenvalue as the measurement outcome. The probability of yielding or collapsing into each eigen- state is given by the absolute square of the projection of the quantum state before the measurement along the eigenstate. The eigenvalue spectrum of an operator can be discrete or continuous or a combination of both. For example, for a one-dimensional infinite square well potential energy, the energy operator $\hat{H}$ (Hamiltonian) has a discrete eigenvalue spectrum, while the position operator $\hat{x}$ has a continuous eigenvalue spectrum. For a generic quantum state, the measurement of energy will collapse the state into one of the energy eigenstates with a well-defined energy, and a measurement of position will collapse the system into an extremely peaked wave function (a position eigenstate) with a well-defined position.

Another reason learning quantum measurement is challenging is that for a given quantum system, the probability of measuring possible values of an observable may change with time depending upon the quantum state and the observable measured [8,50,52,53]. The time development of quantum states is governed by the time-dependent Schrödinger equation (TDSE) $i\hbar\frac{\partial}{\partial t}\Psi(x,t) = \hat{H}\,\Psi(x,t)$. For a time-independent Hamiltonian, by solving the TDSE, the time dependence of a generic state is given by $\Psi(x,t) = e^{\frac{-i\hat{H}t}{\hbar}}\Psi(x,0)$, where $e^{\frac{-i\hat{H}t}{\hbar}}$ is the time-evolution operator. This equation shows that a quantum state at time t is given by acting with the time-evolution operator on the quantum state at time t = 0. If the initial state $\Psi(x,0) = \Psi_n(x)$ is an energy eigenstate, then $e^{\frac{-i\hat{H}t}{\hbar}}\Psi(x,0) = e^{\frac{-i\hat{H}t}{\hbar}}\Psi_n(x) = e^{\frac{-iE_nt}{\hbar}}\Psi_n(x)$, where $E_n$ is the energy corresponding to the energy eigenstate $\Psi_n(x)$. On the other hand, if $\Psi(x,0)$ is not an energy eigenstate, it can be expanded as a linear superposition of energy eigenstates, and the time dependence of this quantum state is given by multiplying each expansion term by a time-dependent phase factor $e^{\frac{-iE_nt}{\hbar}}$, where $E_n$ is the energy corresponding to the $n^{\text{th}}$ energy eigenstate.

As noted, measuring an observable will collapse the state into an eigenstate of the Hermitian operator corresponding to the observable measured. After measuring an observable, the time development of the collapsed quantum state is governed by the TDSE. If the measured observable is energy, the measurement of the observable will collapse the system into an energy eigenstate. Energy eigenstates evolve in time via an overall phase factor, so the probability or probability density of measuring any observable in an energy eigenstate will be time independent. Therefore, energy eigenstates are also called stationary states. On the other hand, if the measured observable corresponds to an operator that does not commute with $\hat{H}$, e.g., position, the collapsed state after the measurement will not be a stationary state. For example, a measurement of position will collapse the system into a position eigenstate, which can be expanded as a linear superposition of energy eigenstates. The time development of the position eigenstate can be obtained by multiplying each expansion term by a corresponding time-dependent phase factor. Since these phase factors are typically different for each term in the expansion, the state will evolve in time in a nontrivial manner, and the probability or probability density of measuring most observables after the position measurement will be time dependent. The only exception is the probability of measuring energy or observables whose corresponding operators commute with $\hat{H}$, since they are constants of motion. This is because the probability of measuring each energy only depends on the modulus of the expansion coefficients when the quantum state is expanded in the energy basis, which are independent of time.

Our prior studies have shown that students have many common difficulties with quantum measurement after lecture-based instruction [46,47,49,54]. For example, some students have difficulties in distinguishing between eigenstates of operators corresponding to different observables [46,47]. Research shows that some students incorrectly think that the eigenstates of any Hermitian operator are stationary states, and some students do not relate the concept of a stationary state with the special nature of the time evolution of that state [46,47]. In addition, some students incorrectly think that an operator acting on a state corresponds to a measurement of the corresponding observable [46,47]. The prior studies also show that many students have difficulties in identifying the probability of measuring different energies in a given state, especially when the quantum state is not explicitly given as a linear superposition of energy eigenstates, and they also have difficulties in identifying the probability density of measuring position given a wave function [46,47]. Some other common difficulties relate to the time development of a quantum state after a measurement [46,47]. For example, prior research shows that some students incorrectly think that a system will stay in a position eigenstate after measuring position, and some others think that a system will evolve back to the initial state (before the measurement) a long time after a measurement of an observable.

In describing and investigating these challenges, we often use the term “student difficulty” to refer to *the use of a specific idea or pattern of reasoning that differs from those we consider correct and appropriate* [55]. Similar to what Heron described [55], in our studies, we aim to uncover what ideas students have after traditional lecture-based instruction that may hinder their ability to develop a good conceptual understanding, as well as what ideas can be built upon to promote student learning through research-based curricula and pedagogies [55]. By posing a variety of conceptual questions across different contexts, we identify both correct and incorrect patterns of reasoning. These insights help curriculum developers and instructors refine instructional approaches to better address students’ conceptual challenges and enhance the effectiveness of physics instruction [55].

One such instructional approach that has proven effective in addressing student difficulties is the use of multiple-choice questions combined with peer instruction [56]. This method was first popularized by Eric Mazur in the physics community [56]. In Mazur’s approach, an instructor poses a multiple-choice question, and students first think by themselves and answer the question anonymously using an electronic response system (clickers) [56]. Then, the students discuss their thoughts with their peers, during which they can compare their answers with each other and explain their reasoning. After the discussion, the instructor may ask for volunteers to share their discussion or give students feedback based on students' performance. In this process, students can have immediate feedback from their peers and instructor, and the instructor can obtain an understanding of students' common difficulties and the percentage of students who understand the concepts. This method using multiple-choice questions with peer instruction has been shown to be effective and relatively easy to incorporate in classes without the need to greatly restructure them [57]. In quantum mechanics, Pollock et al. [25] have developed a set of multiple-choice concept questions that can be embedded within lectures throughout a course, alongside other instructional materials such as homework and tutorials. Our group has also been involved in similar efforts [58].

Ding et al. built on this idea further and developed and validated multiple-choice question sequences (MQS) for introductory physics, which students can respond to in class as a sequence via clickers [59]. A key feature of these MQSs is the careful sequencing of questions, designed to build upon one another, helping learners connect new information with their existing knowledge and progressively deepen their understanding. In our prior studies [52,60–66], we developed and validated MQS for various topics in quantum mechanics, including the Stern-Gerlach experiment [61], time-development of two-state quantum systems [52], and quantum measurement of two-state quantum systems [66], etc. The in-class effectiveness of each validated MQS has been assessed through pre- and post-tests.

While several materials have been developed to enhance student understanding of quantum measurement [25,47,50] and there is some discussion of a preliminary version of a MQS in Ref. [47], no study has specifically addressed the effectiveness of a carefully sequenced set of multiple-choice questions focused on improving student understanding of quantum measurement in systems involving an infinite-dimensional Hilbert space. In this paper, we describe the development, validation, and in-class evaluation of a multiple-choice question sequence for helping students develop a functional understanding of quantum measurement in the context of wave functions in the infinite-dimensional Hilbert space. The theoretical foundation and methodology for this investigation described below are similar to our prior studies [52,60–66].

## II. Theoretical Foundation

The theoretical foundations of our study on improving student understanding of quantum measurement through a research-based multiple-choice question sequence (MQS) are informed by two key frameworks: Preparation for future learning (PFL) and collaborative learning. Below, we introduce these frameworks and explain how they guided the development, validation, and implementation of the MQS.

### A. Preparation for Future Learning

The preparation for future learning (PFL) framework [67,68], developed by Schwartz et al., emphasizes the importance of balancing efficiency and innovation in educational instruction. Efficiency focuses on the rapid acquisition of knowledge, typically through direct instruction, which is common in traditional educational settings. However, while this approach may lead to quick knowledge gains, an overemphasis on efficiency may lead to "routine experts" who struggle to apply their knowledge in novel contexts [67]. On the other hand, innovation encourages exploration and prepares students to apply their knowledge in new situations. Yet, an overemphasis on innovation without sufficient support may lead to "frustrated novices," who are unable to make meaningful progress due to a lack of foundational knowledge or guidance [67]. PFL advocates for an instructional approach that integrates both efficiency and innovation, fostering the development of “adaptive experts” who can use resources effectively and innovatively and transfer their knowledge across different situations [67].

A key concept within the PFL framework is the notion of “a time for telling.” [67] This idea posits that direct instruction, or “telling,” is most effective when it occurs at a specific moment in the learning process—after students have had the opportunity to struggle with innovative material on their own. For instance, Nokes-Malach and Mestre [69] found that students who first struggled with invention tasks outperformed those who only had direct instruction when both groups were later provided with the same resources to explain the concepts. The authors suggest that this initial struggle prepared students for future learning and helped them recognize gaps in their understanding, making them more receptive to subsequent instruction, thereby enhancing their overall learning [69]. In addition, in the context of PFL, “a time for telling” plays a crucial role in balancing innovation with efficiency [67]. While students initially explore and struggle with new concepts (innovation), the strategic use of direct instruction (efficiency) helps solidify their learning and enables them to apply their knowledge in future situations better.

In our study, the PFL framework guides the development, validation, and implementation of the MQS. During the development and validation phase, apart from simpler warm-up questions that are included to get students oriented about relevant concepts, we included questions that most students productively struggle with in individual one-on-one interview situations regardless of whether they ultimately answered the questions correctly. We also included both concrete and abstract questions. While the abstract questions are more innovative, the concrete questions foster efficiency by providing opportunities for students to apply what they have learned in more specific contexts. In the in-class implementation phase, the integration of MQS with lectures as well as the use of collaborative learning and whole-class discussions offer opportunities to balance innovation and efficiency. For example, students initially think about each question and discuss them with peers. This phase fosters productive struggle and innovation by encouraging students to apply their knowledge in novel contexts and creates a “time for telling.” The efficiency phase corresponds to instructors providing targeted instruction and facilitating whole-class discussions to solidify understanding based on student responses to the MQS.

### B. Collaborative learning

The development and implementation of the MQS are also informed by the collaborative learning framework, which emphasizes the importance of social interaction and peer collaboration in the learning process. Schwartz et al. [67,68] highlight that balancing efficiency and innovation in instruction often requires opportunities for students to experiment with ideas and interact with both artifacts and peers, which aligns with the collaborative learning framework.

Collaborative learning, rooted in the social constructivist theory and particularly influenced by Vygotsky’s concept of the zone of proximal development (ZPD) [70], emphasizes the importance of interaction, communication, and social context in the learning process. It suggests that learners can achieve higher levels of understanding and skill when they collaborate with peers or instructors who can provide the necessary support to bridge the gap between what learners can do independently and what they can achieve with assistance [70]. Additionally, the effectiveness of collaborative learning can be further understood through the lens of distributed cognition. Hutchins [71,72] described distributed cognition as the process of sharing cognitive tasks among students, enabling them to build upon each other’s logic, overcome the limitations of individual working memory, and enhance overall learning outcomes. The value of collaborative learning in physics education is well documented [56,73], especially when structures are in place to ensure individual accountability and positive interdependence, such as through grading incentives. For example, research has shown that students working in pairs on conceptual physics questions often outperform those working individually [74,75]. Peer collaboration has also been shown to aid knowledge retention, with studies indicating that pairs of students, even if both initially answered a physics question incorrectly, can arrive at the correct solution approximately 30% of the time through co-construction, and this process appears to aid in better performance on later individual assessments [74,75,76]. Embedded within this collaborative learning framework is the method of peer instruction, popularized by Mazur [56,57], which leverages conceptual multiple-choice questions to stimulate peer discussions. This approach has been associated with improved understanding, performance, and knowledge retention, highlighting the efficacy of collaborative learning in complex domains such as physics.

Collaborative learning framework guides the development, validation and implementation of the MQS sequence in our study. First, the MQS questions were designed and validated to provoke productive discussion among students. Most alternative options or distractors in the MQS incorporate common student difficulties that we identified in prior studies. Although many of these knowledge pieces that students activate are inaccurate [76], they serve as valuable conceptual resources that students bring to the learning process. By including these as distractors, the MQS sequence invites students to discuss and debate different perspectives, encouraging them to critically evaluate their own and others' ideas and refine their thinking through discussion and reflection with their peers. Following single question or sequence of questions that addresses a set of related knowledge points, we include checkpoints and whole-class discussion slides for instructors. These checkpoints are designed to foster peer interaction by giving students the opportunity to organize their knowledge, synthesize interconnected concepts, and collaboratively solidify their understanding. This design reflects the collaborative learning framework's emphasis on structured opportunities for reflection and consensus-building through peer interaction. Additionally, most of the questions focus on the conceptual understanding of quantum measurement rather than rote memorization or procedural tasks, which provokes more in-depth discussion of fundamental principles.

The implementation of MQS is also guided by a collaborative learning framework. Students are first encouraged to think about the questions and tackle them with peers, allowing them to co-construct knowledge through discussion. In this process, distributed cognition plays a key role, as students share cognitive tasks, build upon each other's reasoning, and collectively overcome the limitations of individual working memory. This collaborative effort can help students approach complex problems more effectively by pooling their intellectual resources and refining their ideas through peer interaction. In addition, the whole-class discussion, after receiving proper instruction based on student responses, further can reinforce their learning gains by synthesizing individual and collective insights.

## III. Methodology

### A. Development of the Multiple-Choice Question Sequence

The development of the quantum measurement multiple-choice question sequence (MQS) was inspired by the learning objectives and inquiry-based guided learning sequences of the Quantum Interactive Learning Tutorial (QuILT) focusing on quantum measurement [47]. A QuILT consists of learning sequences that are developed and validated based on cognitive task analysis from both the expert and student perspectives and extensive research on students' common difficulties in learning quantum mechanics. The QuILT uses a guided inquiry-based approach to keep students engaged and build a good knowledge structure. The QuILT should ideally be used in class in small groups but can also be used as a self-paced learning tool, allowing students ample time to engage with the material at their own pace. In our prior studies, we developed, validated, and implemented quantum interactive learning tutorials (QuILTs) to help students learn various quantum mechanics topics including quantum measurement [47], and the implementation of these materials showed encouraging results [19,21,35,36,47,78,79].

The learning objectives of the QuILT focusing on quantum measurement include [47]: 1. Differentiating between a Hermitian operator acting on a state and measurement of an observable. 2. Identifying the possible outcome values and outcome states of quantum measurements and calculating the probability of measuring different values. 3. Describing the time evolution of the quantum system after measuring different observables. 4. Identifying possible outcomes of consecutive quantum measurements. The inquiry-based guided learning sequences in the QuILT typically involved a combination of brief instruction, multiple-choice questions followed by simulations, open-ended questions, and student conversation prompts [47]. The development of multiple-choice questions and then the MQS was inspired by the learning objectives and inquiry-based guided learning sequences of QuILT but involved substantial redesign to tailor the sequence for in-class peer instruction within the time constraints of a typical lecture period.

First, we condensed the extensive content of the QuILT while ensuring that all key concepts were adequately covered. For instance, the QuILT often addressed individual or closely related knowledge points with multiple questions, so in developing the multiple-choice questions which were later sequenced and turned into MQS, we consolidated these questions into fewer, more comprehensive questions. One approach was to incorporate related knowledge points that were separately addressed in multiple QuILT questions into the options of a single multiple-choice question in the MQS. For example, instead of distinct questions for the measurement outcomes of position and energy as in the QuILT, MQS 2.1 presents several statements about possible outcomes of measuring position or energy, asking students to identify all correct statements. These statements include some common incorrect student responses during the implementation of the QuILT. This approach encouraged students to compare and contrast the differences between measuring position and energy while learning collaboratively through discussions, challenging each other's ideas, and refining their understanding.

Second, since the QuILT includes various formats such as open-ended questions and student conversation prompts, we needed to develop multiple-choice questions that could effectively cover these contents. For example, in the QuILT, students were asked to predict the state after time evolution or the outcomes of consecutive measurements, and they were given space to respond to these open-ended questions and then used simulations to check their predictions. In the MQS, we incorporated students' common predictions as options in multiple-choice questions, offering an opportunity for them to critically evaluate their own and others' ideas. Following this, the built-in checkpoint and discussion sections in the MQS guide instructors in using the simulation tool to lead the activity and facilitate a whole-class discussion, helping to address and resolve any inconsistencies in students' understanding.

Third, the multiple-choice questions in the MQS were carefully sequenced to build on one another. For example, section II of the MQS begins with questions in concrete contexts before progressing to the same concepts in more general and abstract contexts. Similarly, in section III, the time evolution of a system after measuring position or energy is first introduced through mathematical expressions and then reinforced with diagrams to solidify students' understanding. Additionally, different concepts may be applied in similar contexts across consecutive questions. For instance, in section IV, students are first asked to identify measurement outcomes without considering time durations in consecutive measurements, followed by questions that include time durations. This careful sequencing enables students to compare and contrast the premises of consecutive questions, deepening their understanding and fostering a more cohesive knowledge structure.

Finally, in the MQS, we added checkpoints and discussion sections following individual questions or sequences of questions that focus on related knowledge. These can be used by instructors to review and emphasize the key concepts covered in the previous questions and to lead general class discussions on broader themes. These checkpoints are designed to give students the opportunity to organize their knowledge frameworks, synthesize interconnected concepts, and collaboratively solidify their understanding.

## B. Development of the pre- and post-test

In addition to the MQS, we developed the corresponding pre- and post-test to assess students' understanding before and after engaging with the MQS. The design of these tests was closely aligned with the learning objectives of the MQS. Compared to the pre- and post-test for the QuILT, the pre- and post-test for MQS have several distinct features. First, the MQS pre- and post-tests include a broader range of consecutive measurement scenarios, such as measuring position followed by position again or energy, and measuring energy followed by energy again or position. For each scenario, we incorporated questions that assess understanding both with and without a time lapse between the measurements, expanding on the QuILT pre- and post-tests, which covered only a subset of these cases. Second, the MQS pre- and post-tests feature a multiple-choice question focused on students' understanding of fundamental concepts related to eigenvalues and eigenstates of energy or position operators. This addition directly corresponds to the learning objectives targeted in the first section of the MQS and was not included in the QuILT pre- and post-tests. Third, the pre- and post-tests for MQS replaced some multiple-choice questions in the QuILT about possible measurement outcomes with open-ended questions. This change was made to elicit richer and more detailed responses from students, providing deeper insights into their understanding. We note that we made two versions of the test, version A and version B. The only difference between these two versions is the given initial state. In version A, the given initial state is $\Psi(x,0) = \sqrt{\frac{2}{7}}\Psi_1(x) + \sqrt{\frac{5}{7}}\Psi_2(x)$, while the given initial state is $\Psi(x,0) = \frac{3}{5}\Psi_1(x) + \frac{4}{5}\Psi_2(x)$ for version B. In the first year of our study, version A was used as pretest and version B was used as post-test, and in the second year of our study, version B was used as pretest and version A was used as post-test. More details about the use of the test will be discussed later in the in-class implementation section.

### C. Validation of the MQS and pre and posttests

The validation of the MQS and the corresponding pretest and post-test was an iterative process involving multiple stages of refinement by PER researchers as well as refinements based on expert and student interviews. This approach ensured that the MQS was pedagogically sound, aligned with the learning objectives, and suitable for use within the time constraints of a typical lecture period.

Using the methods discussed in the development section, we initially developed a preliminary version of the quantum measurement multiple-choice questions which consisted of approximately 25 questions that were later sequenced by researchers into MQS. We then engaged in in-depth discussions with four faculty members, all of whom have rich experience teaching quantum courses, to review the materials and provide feedback. These expert interviews offered valuable insights, ensuring that each question in the MQS and the pre- and post-tests was not only aligned with the learning objectives but also focused on key concepts essential for understanding quantum measurement. Additionally, the experts provided feedback on the clarity of the question wording and the logical flow of the sequence, ensuring that the questions were clear, unambiguous, and sequenced in a way that allowed students to progress smoothly from one concept to the next. Given the time constraints of in-class instruction, the experts also identified which concepts could be combined or condensed without sacrificing educational value, allowing for more efficient coverage of the material. We note that throughout the refinement process, undergraduate students in the quantum mechanics course were also involved in testing the questions and the feedback collected from them also informed the iterative revisions of the MQS and pre- and post-tests.

After several rounds of expert feedback and revision, we developed a version of the quantum measurement MQS consisting of 16 questions. Then, we conducted individual interviews with 10 students, totaling approximately 20 h using this version. During these sessions, students completed the pretest, followed by the MQS, and then the posttest. We employed a think-aloud protocol, in which students verbalized their thoughts while answering each question. This approach enabled us to observe how students navigated through the MQS and the challenges they encountered. We did not disturb them when they thought aloud to not disrupt their thought processes. After each MQS, we first asked students for clarification of the points they may not have made, and then we led discussions with them on each choice in the MQS question as appropriate and their perceptions of how each question relates to others. These discussions provided further insights into how the sequence of questions impacted students' processing of information, allowing us to refine the structure of MQS for a better flow. The interviews also offered valuable data on the difficulty level of each question, guiding us in making adjustments to ensure an appropriate balance of difficulty. We note that some questions are designed as warm-up questions and serve as an initial step in the sequence to get students' thoughts organized. While these questions may have relatively low difficulty, they are still valuable. Other questions are specifically intended to promote productive struggle and deepen student understanding. For these questions, if most students arrived at the correct answer with correct reasoning (correctness at around 70% or above), we added some challenging options to create more opportunities for productive struggle. Conversely, for questions that proved very challenging for most students in the interviews (correctness at round 30% or below), we incorporated additional scaffolding within the questions or in preceding questions to better support student understanding. However, as the sample size of the interviewed students is small, we did not have strict criteria for easy and difficult questions (particularly for in-class implementation) but we wanted most questions in the sequence to be such that students would have productive discussions with each other in class. During the think-aloud sessions for the pre- and post-tests, we also paid attention to whether students interpreted the questions as intended and whether their reasoning aligned with their approach during the MQS, and whether the pre- and post-tests sufficiently covered all learning objectives discussed earlier. Based on the feedback from both the expert and student interviews, we iteratively adjusted the MQS and the pre- and post-tests, refining the materials over several iterations to ensure their clarity and effectiveness in facilitating student learning. These student interviews revealed common difficulties in understanding quantum measurements, consistent with our previous studies [46,47]. The interviews also showed that after working through the whole MQS, students' difficulties with many concepts related to quantum measurement were reduced. The interviewed students also reported that they found the scaffolding provided by the sequenced questions and discussion slides helpful.

### D. Structure of the Quantum Measurement MQS and the pre- and post-test

The quantum measurement MQS in its final iteration includes four sections (see Appendix A), with each section focusing on a broad learning objective discussed earlier. The first section of the MQS includes two questions (MQS 1.1 and 1.2). The learning objective of this section is to help students differentiate between an operator acting on a state and the measurement of an observable. The second section of the MQS includes three questions (MQS 2.1-2.3), which help students learn about the postulates related to quantum measurement, e.g., a measurement of an observable collapses the system into an eigenstate of the operator corresponding to that observable and returns the eigenvalue corresponding to the eigenstate. In particular, the learning objectives include being able to identify the possible outcomes of the measurement of an observable and calculate the probability (for observables corresponding to operators with discrete eigenvalue spectra) or probability density (for observables corresponding to operators with continuous eigenvalue spectra) of measuring outcomes for a given quantum state. MQS 2.2 helps students recognize that an operator acting on a state is not equivalent to a measurement of the corresponding observable in that state, which has been shown to be a common difficulty in our prior studies [46,47]. The third section of the MQS includes six questions (MQS 3.1–3.6), which aim to help students learn the time development of the quantum state after a measurement of an observable. In particular, MQS 3.1 and 3.2 help students to learn about the time development of a generic state and a stationary state (energy eigenstate) and the fact that one can always expand a generic state as a linear superposition of energy eigenstates. In MQS 3.3–3.5, we use both mathematical and pictorial representations to help students learn about how a quantum system evolves in time after the measurement of energy or position, respectively. MQS 3.6 helps students learn about whether the probability (for operators with discrete eigenvalue spectra) or probability density (for operators with continuous eigenvalue spectra) for measuring an observable depends on time. This question also prepares students for the next section of the MQS, pertaining to consecutive measurement. In the last section (MQS 4.1–4.5), MQS 4.1 and 4.2 help students identify the possible outcomes and the corresponding probability densities of measuring position immediately after or a long time after a measurement of energy. MQS 4.3 and 4.4 help students identify the possible outcomes and the corresponding probability densities of measuring position immediately after or a long time after a measurement of position. MQS 4.5 helps students identify the possible outcomes and the corresponding probabilities of measuring energy immediately after or a long time after a measurement of position.

The final version of the pretest and post-test (see Appendix B) contains four questions. We note that Appendix B only shows the version A of the test. The only difference between these two versions is the given initial state. In version A, the initial state is $\Psi(\mathrm{x},0)=\sqrt{\frac{2}{7}}\Psi_1(\mathrm{x})+\sqrt{\frac{5}{7}}\Psi_2(\mathrm{x})$, while the initial state is $\Psi(\mathrm{x},0)=\frac{3}{5}\Psi_1(\mathrm{x})+\frac{4}{5}\Psi_2(\mathrm{x})$ for version B. The first question is a multiple-choice question, while the remaining questions are open-ended, requiring both an answer and corresponding reasoning. Question 2 is divided into two sub-questions (2a, 2b), and Questions 3 and 4 each include five sub-questions (3a–3e, 4a–4e). Question 1.1 evaluates whether students understand that the action of the Hamiltonian operator acting on a generic state does not correspond to the measurement of energy and leads to the collapse of the state. Similarly, question 1.2 evaluates student understanding of the position operator acting on a generic state. Questions 2a and 2b assess students' understanding of the possible outcomes and the corresponding probability or probability density of measuring energy or position in a given quantum state. Questions 3 and 4 evaluate students' understanding of the time evolution of a quantum system after measuring energy or position and the possible outcomes and corresponding probability or probability densities of measuring energy or position in a consecutive sequence.

### E. Participants and course context

The participants in this study were students enrolled in a junior- or senior-level quantum mechanics course at a large research university in the United States, for which calculus 1–3 and linear algebra are prerequisites. The study was conducted over two consecutive academic years, with the same instructor teaching the course in both years. In the first year, 23 students participated in the pretest, and 25 students completed the MQS and post-test. In the second year, 18 students participated in both the pre- and post-test. To compare students' performance in the pre- and post-test, we focused on those who completed both the pre- and post-test (N = 41). All participants were physics majors. The instructor used Griffiths' *Introduction to Quantum Mechanics* [51] as the primary textbook but followed a "spins-first" approach. The course began with the chapter on formalism (chapter 3), followed by spin-1/2 in chapter 4, before returning to chapters 1 and 2 on wave functions in infinite-dimensional Hilbert space. This approach allowed the course to cover material related to two-state systems early, before introducing "infinite square well" and "simple harmonic oscillator" involving infinite-dimensional Hilbert space. The course meets for three 50-min class periods each week. Students were assigned weekly homework, and the course included two midterm exams and a final exam. The instructor has significant experience teaching quantum mechanics at the undergraduate level. Additionally, the instructor is highly supportive of physics education research and has previously utilized research-based learning tools in their teaching.

### F. In-Class Implementation

Before the in-class implementation of the MQS, students had traditional lecture-based instruction covering all concepts related to quantum measurement targeted in the pretest. Around 5.5 class periods over several weeks, prior to the pretest, were dedicated to the topic of quantum measurement, distributed throughout the semester. Specifically, when introducing the formalism of quantum mechanics in chapter 3, about 2 class periods were dedicated to lectures on the postulates of quantum measurement without reference to a specific quantum system such as an infinite square well or simple harmonic oscillator consistent with the treatment in the Griffiths' textbook. When introducing chapter 4, around 1.5 class period focused on quantum measurement in the context of spin1/2 systems. Similarly, around two class periods were dedicated to quantum measurement in the context of a one-dimensional infinite square well and harmonic oscillator during the introduction of chapter 2. Corresponding homework assignments were given spread over different weeks as these topics were covered in the lectures from different chapters. Thus, before taking the pretest, students had been exposed to the topic of quantum measurement multiple times throughout the semester in various contexts including when the instructor covered quantum measurement in chapters 3, 4, and 2. The pretest was administered during the subsequent class period following the lecture, taking about 25 min. The MQS was then implemented over the next two consecutive class periods, each lasting 50 minutes. After completing the MQS, the post-test was administered during the subsequent class period, again taking about 25 min.

During the MQS sessions, students were first asked to think about each MQS question and discuss their thoughts with their peers before voting using electronic response systems (clickers) [56]. The time given to students for thinking and answering each question is around 1–2 min. If most students answered correctly (e.g., a warm-up question), the instructor would briefly review the question, emphasize key concepts, address any confusion, and then move on to the next question. For questions that are more challenging, it took longer for students to discuss and respond. If a significant number of students answered incorrectly, the instructor provided additional scaffolding, encouraged further discussion, and sometimes had students revote. After a single question or sequence of questions that addresses a set of related knowledge points, the instructor would lead a whole-class discussion to explore the relationships between the concepts covered in the previous questions and to check students' understanding. On average, around 4–5 min was devoted to class discussion for each question. Therefore, the time ratio devoted to MQS and peer instruction to general class discussions is around 2/5.

The only other content assigned during this period was a traditional textbook homework from chapter 2, assigned in the same week. While this study did not include a control group that completed only the homework for comparison, our previous research has shown that student post-test performance significantly improves when research-based tools like the MQS are used in instruction, compared to control groups of students who receive only traditional instruction and homework [47]. Also, in this study, students had done some homework problems on quantum measurement from chapters 3 and 4 before the pretest. That said, this study is nonetheless quasi-experimental in design, given the various factors such as homework, over which we did not have complete control.

As noted earlier, there are two versions of the test to evaluate student learning gains. In the first year of our study, version A was used as the pretest and version B as the post-test, while in the second year, version B was used as the pretest and version A as the post-test. Students' responses to the pretest and post-test were graded by two researchers. Each sub-question was scored out of 1 or 2 points, with partial credit given for correct answers that lacked proper justification or provided incorrect reasoning when justification was requested. The inter-rater reliability was greater than 95%.

## IV. Results and Discussion

Table 2 compares students' performances on the pretest (after traditional lecture-based instruction) and on the posttest (after students had engaged with the quantum measurement MQS) across 2 years of implementation of the MQS. The normalized gain was calculated as $g$ = (post% – pre%)/(100% – pre%) [80,81]. The effect size was calculated as Cohen's $d = (\mu_{post} - \mu_{pre})/\sigma_{pooled}$, where $\mu_{pre}$ and $\mu_{post}$ represent students' average correctness in the pre-test and post-test, respectively, and $\sigma_{pooled}$ is the pooled standard deviation, which is the weighted average of standard deviations of pre- and post-test [80]. Both measures offer valuable insights into student performance but have different strengths and limitations. Normalized gain is effective at capturing relative improvement in student performance, making it useful for assessing how much progress has been made relative to what was possible. However, it is sensitive at the extremes; small improvements near the maximum score and large gains from lower baselines can yield similar normalized gains, despite representing different educational outcomes [82]. On the other hand, Cohen's d represents the standardized difference between two means, providing a more stable measure of the magnitude of change, independent of initial performance levels [82].

In this study, we report both normalized gain and Cohen's d together to provide complementary perspectives of student learning outcomes, and we classified the degree of student improvement from the pretest to the post-test based on Cohen's $d$: improvements with Cohen's $d \geqslant 0.70$ were categorized as "major improvement," while improvements with Cohen's d $< 0.70$ were categorized as "some improvement." As shown in Table 1, most questions show improvements across both years of implementation. In particular, students in both years exhibited major improvements in questions 4a, 4b, and 4d. In addition, year 1 students showed major improvements in questions 2b and 4c, while year 2 students showed major improvements in questions 3c and 4e.

We note that students in year 2 began with higher pretest scores on some questions compared to those in year 1. As mentioned earlier, the only difference between the pretests administered in the 2 years was the coefficients in the given initial state, which is unlikely to result in a significant difference in perceived difficulty for students. Thus, this difference in initial performance may indicate that instructor's traditional instruction may have improved after the first year of MQS implementation, as they became more familiar with the common difficulties students face in understanding quantum measurement so they may have paid more attention to these difficulties during the lecture. However, in the following analysis in Table 2, we combined the data from both years to balance out potential biases and to increase the sample size, thereby enhancing the statistical power of our analysis.

**Table 1.** Comparison of mean pre-test and post-test scores for each question, along with corresponding normalized gains and effect sizes, for students who engaged with the quantum measurement MQS (In Year 1, N=23 for both pre- and post-tests; in Year 2, N=18 for both pre- and post-tests).

| Question | Year 1 | | Year 2 | | Normalized gain | | Cohen's *d* | |
|---|---|---|---|---|---|---|---|---|
| | Pre (version A) | Post (version B) | Pre (version B) | Post (version A) | Year 1 | Year 2 | Year 1 | Year 2 |
| 1 | 81% | 84% | 81% | 89% | 0.15 | 0.43 | 0.12 | 0.32 |
| 2a | 93% | 98% | 89% | 100% | 0.67 | 1.00 | 0.24 | 0.57 |
| 2b | 43% | 85% | 69% | 89% | 0.73 | 0.64 | 1.18 | 0.58 |
| 3a | 86% | 93% | 94% | 94% | 0.50 | 0.00 | 0.25 | 0.00 |
| 3b | 91% | 91% | 94% | 94% | 0.00 | 0.00 | 0.00 | 0.00 |
| 3c | 63% | 72% | 63% | 93% | 0.24 | 0.81 | 0.19 | 0.83 |
| 3d | 57% | 70% | 61% | 81% | 0.30 | 0.50 | 0.33 | 0.54 |
| 3e | 48% | 67% | 58% | 78% | 0.38 | 0.47 | 0.42 | 0.42 |
| 4a | 13% | 65% | 39% | 89% | 0.60 | 0.82 | 1.24 | 1.18 |
| 4b | 65% | 91% | 64% | 94% | 0.75 | 0.85 | 0.73 | 0.81 |
| 4c | 48% | 93% | 86% | 100% | 0.88 | 1.00 | 1.20 | 0.59 |
| 4d | 0% | 41% | 14% | 75% | 0.41 | 0.65 | 1.19 | 1.38 |
| 4e | 33% | 59% | 31% | 61% | 0.39 | 0.44 | 0.63 | 0.77 |

**Table 2.** Comparison of mean pre-test (weighted average of versions A and B) and post-test scores (weighted average of versions B and A) for each question, along with corresponding normalized gains and effect sizes, for students who engaged with the quantum measurement MQS using combined data from two years of implementation (N=41 for both the pre-test and post-test).

| Question | Pre-test mean | Post-test mean | Normalized gain | Cohen's *d* |
|---|---|---|---|---|
| 1 | 81% | 86% | 0.28 | 0.21 |
| 2a | 91% | 99% | 0.86 | 0.40 |
| 2b | 55% | 87% | 0.70 | 0.90 |
| 3a | 89% | 93% | 0.38 | 0.16 |
| 3b | 93% | 93% | 0.00 | 0.00 |
| 3c | 63% | 81% | 0.49 | 0.43 |
| 3d | 58% | 74% | 0.38 | 0.42 |
| 3e | 52% | 72% | 0.41 | 0.42 |
| 4a | 24% | 76% | 0.68 | 1.18 |
| 4b | 65% | 93% | 0.79 | 0.78 |
| 4c | 65% | 96% | 0.90 | 0.91 |
| 4d | 6% | 54% | 0.51 | 1.24 |
| 4e | 32% | 60% | 0.41 | 0.70 |

Table 2 shows the mean pre-test and mean post-test scores on each question with the combined two years of data. As shown in the table, students' average correctness for questions 1, 2a, 3a, and 3b in the pretest was higher than 80%, which indicates that after a traditional lecture-based introduction, students had a relatively good understanding of single energy measurement and consecutive energy measurements. This result is reasonable, considering there were approximately 5–6 class periods of lecture instruction on quantum measurement topics, along with homework assignments on quantum measurement from chapters 3 and 4, prior to the pretest. However, the average correctness for other questions was low in the pretest. In particular, students' average correctness on questions 4a, 4d, and 4e was below 40% in the pre-test. Question 4a tests students' understanding of the wave function immediately after a position measurement, which is a delta function, and questions 4d and 4e assess students' understanding of the possible outcomes of an energy measurement immediately after or a long time after the measurement of position in 4a. Students' average correctness for the rest of the questions (2b, 3c, 3d, 3e, 4b, and 4c) was around 60% on the pretest, which indicates that students also had difficulties in understanding the concepts assessed by these questions. In particular, question 2b assesses students' understanding of the possible outcomes and the corresponding probability densities of measuring position in a given quantum state. Questions 3c, 3d, and 3e assess students' understanding of the possible outcomes of energy or position measurements after a measurement of energy. Questions 4b and 4c assess students' understanding of the possible outcomes of a position measurement made immediately after or a long time after a measurement of position.

By comparing students' average correctness on different questions, we observed that students were less likely to answer correctly on questions involving position measurement (such as 4b) compared to those involving energy measurement (such as 3b). Students also struggled more with questions that required identifying the possible outcomes of consecutive measurements (such as 3c) than with those involving a single measurement (such as 2a and 3a). Additionally, students' average correctness on questions involving consecutive measurements with a time interval between them (such as 3c) was lower than on those involving measurements made in immediate succession (such as 3b). Moreover, questions about measuring energy after a position measurement (such as 4d and 4e) appeared to be more challenging for students compared to questions about measuring position after an energy measurement (such as 3d and 3e). Below, we provide a detailed analysis of the specific areas where students struggled and compare their performance on the pretest and post-test.

Table 2 shows an overall improvement in students' performance from the pretest to post-test. In particular, questions 2b, 4b, and 4c show major improvements. We note that the average correctness was around 60% for questions 2b, 4b, and 4c on the pretest, and it improved to around 90% on the post-test. Question 2b asks about the probability density for measuring position in state $\Psi(x,0) = \sqrt{\frac{2}{7}}\Psi_1(x) + \sqrt{\frac{5}{7}}\Psi_2(x)$ for one-dimensional infinite square well (0<x<a). One common difficulty we found from student-written responses is that some students could not distinguish between the eigenstates of energy and position operators [46,47]. Some students wrote that the probability density of measuring position in the state $\Psi(x,0)$ is "$|\Psi_1(x)|^2$ or $|\Psi_2(x)|^2$". One possible reason for this difficulty could be that since wave functions are often discussed in the context of energy eigenstates or written as a linear superposition of energy eigenstates, students may assume that each term in the linear superposition corresponds to a possible measurement outcome regardless of the observable measured. Some students wrote the possible outcomes of a position measurement as "$x_n$". Another possible explanation is that students may have assumed that the energy was measured first (as in question 2a) before the position measurement.

Questions 4b and 4c assess students' understanding of the possible outcomes of a position measurement made immediately after or a long time after the first measurement of position. We note that on the pretest, some students wrote that the second measurement of the position will always yield the same result as the first position measurement regardless of whether there is a time interval between the two measurements because "the first measurement collapsed the state" or "the position measurement is time independent." These types of answers reflect students' difficulties in distinguishing between stationary states and eigenstates of other operators [46,47], e.g., position, that do not commute with the Hamiltonian. Some other students noted that position eigenstates evolve with time, but they thought that the time development of position eigenstates is such that the probability density of measuring position is not affected. For example, one student who stated that the probability of measuring position is time independent explained, "the time [phase] factor will be all that changed." This reasoning also indicates that students may have difficulty in distinguishing between the time development of energy eigenstates and those of other operators such as position [46,47]. As shown in Table 2, after the implementation of the MQS, students had significant improvement in their performance related to single position measurement and consecutive position measurements (evidenced by performance on questions 2b, 4b, and 4c).

As shown in Table 2, while questions 4a, 4d, and 4e also show major improvements based on our criteria, the average correctness percentages for questions 4a, 4d, and 4e were very low on the pretest and there is still room for improvement in the post-test. Question 4a asks students to identify the wave function after a measurement of position in state $\Psi(x,0) = \sqrt{\frac{2}{7}}\Psi_1(x) + \sqrt{\frac{5}{7}}\Psi_2(x)$ with outcome $x_0$, and questions 4d and 4e ask students to identify the possible outcomes of an energy measurement made immediately after or a long time after the position measurement in 4a. As we can see in Table II, the correctness for question 4a was only 24% in the pretest. The most common student response was that the wave function right after the position measurement is $\sqrt{\frac{2}{7}}\Psi_1(x_0) + \sqrt{\frac{5}{7}}\Psi_2(x_0)$, which is obtained by simply replacing the variable $x$ in the initial state $\Psi(x,0)$ with position $x_0$. However, this expression is actually the probability amplitude at $x = x_0$ before the position measurement. This result shows that some students had difficulties in recognizing that a position measurement will instantaneously collapse the wave function to a delta function in position [46]. Another common incorrect answer to question 4a was "$\Psi_1(x_0)$ or $\Psi_2(x_0)$", which also shows that some students had difficulties in distinguishing between energy eigenstates and position eigenstates as discussed earlier [46,47]. After the implementation of the MQS, around two-thirds of the students who answered 4a incorrectly in the pretest were able to recognize that a delta function is the outcome for the state right after a position measurement in the post-test, while some students still answered "$\Psi(x_0)$" or "$\Psi_1(x_0)$ or $\Psi_2(x_0)$".

Question 4d asks students to identify the probability of measuring $E_1$ immediately after the position measurement in 4a. This question requires students to expand the delta function as a linear superposition of energy eigenstates and then express the probability of measuring $E_1$ using the expansion coefficient for the ground state. We note that on the pretest, no students provided the correct response. Many students wrote that the probability of measuring $E_1$ is $\frac{2}{7}$, obtained by simply squaring the coefficient of $\Psi_1(x)$ in the initial state $\Psi(x,0)$. This response suggests that these students failed to recognize that after a position measurement, the wave function collapses to a position eigenstate. Since the initial wave function was not a position eigenstate, measuring position would change the wave function to a new function, which is a different superposition of energy eigenstates. In other words, these students appeared to ignore the collapse of the wave function during the position measurement, leading them to incorrectly use the initial state's energy expansion coefficients to calculate the probability of measuring $E_1$. Table 2 shows that on the post-test, many students were able to express the probability of measuring $E_1$ by projecting the delta function along state $\Psi_1(x)$. For example, on the post-test, some students wrote statements such as "the system is in a superposition of an infinite number of energy eigenstates and the probability of measuring $E_1$ is $|\langle\Psi_1(x)|\delta(x-x_0)\rangle|^2$". Even though keeping the wavefunction with $x$ in Dirac notation is not accurate, this example shows students' improvement at the conceptual level about probability of measuring energy after the implementation of the MQS.

Question 4e asks about the time dependence of the probability of measuring E1 after a position measurement. This question is challenging for students because the system itself is not in a stationary state and will evolve in time in a nontrivial manner, but the probability of measuring each eigenenergy is time independent since the Hamiltonian is a constant of motion. Students need to recognize that any state can be expanded as a linear superposition of energy eigenstates, and each expansion coefficient evolves in time via a different phase factor, but the phase factor will cancel out when calculating the probability of measuring a specific energy (by taking the absolute square). A common difficulty that students had was that they thought that since the state is not a stationary state, the probability for measurement of a particular value of energy should also change with time [46,47]. Moreover, some students stated that the probability of measuring $E_1$ will be $\frac{2}{7}$ with explanation such as this, "after a long time, the wavefunction will evolve back to $\Psi(x,0) = \sqrt{\frac{2}{7}}\Psi_1(x) + \sqrt{\frac{5}{7}}\Psi_2(x)$". On the post-test, the correctness of 4e increased indicating that students' understanding of energy measurements after a position measurement improved after the implementation of the quantum measurement MQS.

In addition, Table 2 shows some improvement in students' performances on questions 3c, 3d, and 3e. The correctness of these questions on the pretest was around 50%–60% and improved to around 70%–80% on the posttest. These three questions ask about the possible outcomes of measuring energy or position immediately after or a long time after a measurement of energy. Question 3c asks about the possible outcomes of measuring energy a long time after an energy measurement that yields $E_1$. One common difficulty with this question was that some students thought that the system would evolve to a linear superposition of energy eigenstates after the first measurement of energy. For example, one student stated that the second energy measurement would yield a different value from the first one because "the wave function will no longer be collapsed, so you could measure $E_1$ or $E_2$ [for the second measurement]". Question 3d asks about the probability density of measuring position at $x = x_0$ immediately after an energy measurement that yields $E_1$. Even though many students correctly identified that after the energy measurement, the state will collapse to $\Psi_1(x)$, some of them still had difficulties in identifying the probability density for the position measurement [46,47]. For example, one student stated, "the probability density is 100% at $x = x_0$ because the system has collapsed to $\Psi_1(x)$". This response again shows students' difficulty in distinguishing between eigenstates of different operators corresponding to observables (e.g., Hamiltonian and position).

In addition, we note that some students had difficulties in writing the probability density for a given wave function, which was reflected in students' responses to both questions 2d and 3d. For example, some students stated something similar to this student's explanation, "the probability density of measuring position is 0 because position is a single point," and some put a position operator in front of the wave function, e.g., "$|\hat{x}\Psi_1(x)|^2$", or multiplied the wavefunction by $x$, e.g., "$|x\Psi_1(x)|^2$" to represent the probability density. Some students had difficulties in distinguishing between probability density and probability and used an integral such as $\int_0^{x_0}|\Psi_1(x)|^2dx$ to represent probability density. Question 3e asks whether the probability density of measuring position changes with time after an energy measurement. Students who incorrectly stated that the probability density will change with time usually had two types of reasoning: the state itself will evolve with time after the energy measurement or since the position operator does not commute with the energy operator (Hamiltonian), the measurement of position is time dependent even in an energy eigenstate. In the post-test, many students correctly stated that for a stationary state, the probability or probability density of measuring any observable will be time independent.

Even though most students answered questions 1, 2a, 3a, and 3b correctly on the pretest as shown in Table II, we find some common difficulties among students who did not answer these questions correctly. For example, in our interviews, students who chose $\hat{H}\Psi(x) = E_n\psi_n(x)$ as a correct statement in question 1 explained that acting with an operator on a state will collapse this state to an eigenstate of the operator corresponding to the observable, which is consistent with the findings of prior studies [46,47]. We also found that some students were confused about the normalization of quantum states after energy measurement. For example, question 3a asks about the normalized state after a measurement of energy in state $\Psi(x,0) = \sqrt{\frac{2}{7}}\Psi_1(x) + \sqrt{\frac{5}{7}}\Psi_2(x)$ that yields $E_1$, and some students incorrectly stated $\sqrt{\frac{2}{7}}\Psi_1(x)$. A summary of the student difficulties found in written pre-test and post-test is presented in Table 3.

**Table 3. Summary of the learning objectives and related conceptual difficulties addressed by the MQS. Below, specific examples of difficulties with quantum measurement along with the MQS questions that address them are listed. In the comments section, we include the relevant pre-/post-test question numbers and whether post-test showed 'some' improvement or 'major' improvement compared to the pre-test. Improvements with Cohen's $d \geq 0.70$ were categorized as "Major improvement," while improvements with Cohen's $d < 0.70$ were categorized as "Some improvement".**

| Learning objectives | **MQS** | Difficulties | Test questions and comments |
|---|---|---|---|
| Differentiating between operator acting on a state and measurement of an observable. | 1.1, 1.2, 2.2 | Responses consistent with incorrect reasoning that an operator acting on a state corresponds to a measurement of the corresponding observable. | 1<br>Some improvement |
| Identifying the possible outcome values and outcome states of measurements. | 2.1, 2.2, 3.3, 3.4 | **Position measurement**<br>• Stating that a position measurement in a given state will yield discrete possible outcomes, e.g., "$x_n$".<br>• Not recognizing that the wavefunction will collapse to a delta function after a position measurement:<br>  • Stating that the wavefunction after a position measurement that yields $x_0$ is $\sqrt{\frac{2}{7}}\Psi_1(x_0) + \sqrt{\frac{5}{7}}\Psi_2(x_0)$<br>• Stating that the wavefunction after a position measurement that yields $x_0$ is $\Psi_1(x_0)$ or $\Psi_2(x_0)$. | 2b<br>Major improvement<br><br>4a<br>Major improvement |
| | 2.1, 2.2, 3.3, 3.5 | **Energy measurement**<br>• Not writing the correct normalized state after an energy measurement, e.g., stating that the normalized state after a measurement of energy in state $\sqrt{\frac{2}{7}}\Psi_1(x)+\sqrt{\frac{5}{7}}\Psi_2(x)$ that yields $E_1$ is $\sqrt{\frac{2}{7}}\Psi_1(x)$.<br>• Difficulties in distinguishing between measurement outcomes and probability amplitude of measuring them, e.g., stating that the possible values obtained from an energy measurement in state $\sqrt{\frac{2}{7}}\Psi_1(x)+\sqrt{\frac{5}{7}}\Psi_2(x)$ are $\sqrt{\frac{2}{7}}$ and $\sqrt{\frac{5}{7}}$. | 2a<br>Some improvement<br><br>3a<br>Some improvement |
| Calculating the probability or probability density of measuring energy and position. | 2.3, 3.6, 4.1, 4.2, 4.4 | **Position measurement**<br>• Stating that the probability density for measuring position in state $\sqrt{\frac{2}{7}}\Psi_1(x)+\sqrt{\frac{5}{7}}\Psi_2(x)$ is "$\|\Psi_1(x)\|^2$ or $\|\Psi_2(x)\|^2$".<br>• Stating that the probability density of measuring position in state $\sqrt{\frac{2}{7}}\Psi_1(x)+\sqrt{\frac{5}{7}}\Psi_2(x)$ is e.g., 2/7 or 5/7.<br>• Using an incorrect integral to represent probability density, e.g., "$\int \hat{x}\Psi_1(x)dx$" or "$\int_0^a x_0\|\Psi_1(x_0)\|^2dx$".<br>• Acting with the position operator on the wavefunction, e.g., "$\|\hat{x}\Psi_1(x)\|^2$" or multiplying the wavefunction by $x$, e.g., "$\|x\Psi_1(x)\|^2$" to represent probability density.<br>• Stating that the probability density is 0 because position is a single point. | 2b<br>Major improvement<br><br>3d<br>Some improvement |
| | 2.3, 3.1, 3.6, 4.5 | **Energy measurement**<br>• Not recognizing the need to expand a generic wavefunction as a linear superposition of energy eigenstates to identify the probability of measuring different energies.<br>• Difficulties in calculating the expansion coefficients for each energy eigenstate when a generic wavefunction is not explicitly written as a linear superposition of energy eigenstates. | 4d<br>Major improvement<br><br>4e<br>Major improvement |
| Describing the time evolution of the quantum system after measuring different observables. | 3.1, 3.2, 3.3, 3.5 | **Stationary state**<br>• Not recognizing that energy eigenstates evolve in time via a trivial overall time-dependent phase factor.<br>• Stating that a quantum system will evolve back to its original state (before the measurement) a long time after a measurement of energy. | 3c<br>Some improvement<br><br>3e<br>Some improvement |
| | 3.1, 3.3, 3.4, 3.6 | **Non-stationary state**<br>• Stating that a position eigenstate does not evolve in time (system is stuck in a position eigenstate because it is an eigenstate). | 4c<br>Major improvement<br><br>4e<br>Major improvement |
| | | • Stating that a quantum system will evolve back to its original state (before the measurement) a long time after a measurement of position.<br>• Not recognizing that the probability density for position measurement in a non-stationary state changes with time. | |
| Identifying possible outcomes and probability after consecutive measurements. | 3.3, 3.5, 4.2 | **Measurement after measuring energy**<br>• Not recognizing that the probability/probability density of measuring any observable (with no explicit time-dependence) is time independent in an energy eigenstate (stationary state). | 3c<br>Some improvement<br><br>3e<br>Some improvement |
| | 3.3, 3.4, 3.6, 4.3, 4.4, 4.5 | **Measurement after measuring position**<br>• Not recognizing that consecutive position measurements made in immediate succession will yield the same outcomes.<br>• Stating that a position eigenstate evolves in time via a trivial overall time-dependent phase factor so that the probability density for position measurement will be time-independent if the system was in a position eigenstate at time $t = 0$<br>• Not recognizing that the probability of measuring a particular value of energy is time independent regardless of state since energy is a constant of motion.<br>• Stating that the probability of measuring ground state energy after a position measurement is 100% because the state has collapsed. | 4b<br>Major improvement<br><br>4c<br>Major improvement<br><br>4d<br>Major improvement<br><br>4e<br>Major improvement |

## a. Summary and general discussion

In this study, we discussed the development, validation, and in-class implementation of a multiple-choice question sequence (MQS) designed to enhance students' understanding of quantum measurements for quantum systems in an infinite-dimensional Hilbert space. The MQS was developed based on the learning objectives and inquiry-based sequences from the interactive tutorial (QuILT) that we previously developed for the same topic [46,47]. It condenses the extensive content of the QuILT into a format that can be effectively administered within the constraints of limited class time. Additionally, the multiple-choice options include common incorrect student responses as distractors, providing opportunities for peer instruction and productive struggle. Furthermore, the questions in the MQS were carefully sequenced to build upon each other, helping students organize, extend, and refine their knowledge structure while developing a deeper understanding of various aspects of quantum measurement. The 16 multiple-choice questions in the MQS are divided into four sections, each focusing on one broad learning objective.

The quantum measurement MQS was implemented in a junior-/senior-level quantum mechanics course at a large research university in the U.S. The implementation involved students answering the questions anonymously using an electronic response system (clickers) [83], followed by peer instruction and feedback from the instructor. We used a pre-/post-test to assess students' understanding before and after the implementation of the MQS. We found that even after traditional lectures, students continued to struggle with several concepts related to quantum measurement, such as distinguishing between applying an operator to a state and measuring an observable in a state, differentiating between energy eigenstates and eigenstates of other operators, and understanding the time evolution of a quantum system after a measurement. These findings are consistent with prior studies on student understanding of quantum measurement [46,50,84].

Our results show that the average correctness across questions improved after the MQS. The most significant improvements were observed in students' performance on questions related to single position measurement, consecutive position measurements, and energy measurements following a position measurement. There were also improvements in questions related to measurements of energy or position taken immediately after or a long time after an energy measurement. While prior studies [46,50] have demonstrated that research-based tutorials can enhance students' understanding of quantum measurement, these tutorials often involve open-ended questions and typically require a substantial time investment, such as during class, recitation sessions, or as homework assignments. This study contributes to the literature by showing that a carefully sequenced multiple-choice question set can effectively enhance student understanding of quantum measurements in systems involving infinite-dimensional Hilbert spaces.

We note that our results eliminate the possibility that students were learning through the test itself for several reasons. First, the later questions on the post-test, such as 4d and 4e (which involve measuring energy after a position measurement), were more challenging for students than the earlier questions, like 3d and 3e (which involve measuring position after an energy measurement). This suggests that students were not simply applying knowledge gained from earlier questions to later ones. Moreover, even though both question sets 3 and 4 focus on consecutive measurements, they differ significantly. In question 3, the measurement sequence begins with energy measurement, while in question 4, it begins with position measurement. The subsequent time development of the system after energy measurement and position measurement is very different, making it unlikely that students could learn how to approach question 4 based on their experience with question 3 alone.

In addition, we note that among the 41 students matched from pretest to post-test, only three did not show improvement. Two of these students continued to struggle with understanding the time evolution of a quantum system after measuring energy, compared to after measuring other observables. The third student still had difficulty correctly expressing the probability density for position measurements. These challenges are reflected in their responses to questions 3e and 3d, which some other students also continued to find difficult. Additionally, in question 4a, some students still struggled with the fact that the wave function collapses to a delta function after a position measurement. For questions 4d and 4e, while some students did not answer correctly on both the pre- and post-tests, there were noticeable shifts in their reasoning. For instance, in question 4d, which asks for the probability of measuring $E_1$ immediately after a position measurement of an initial state $\sqrt{\frac{2}{7}}\Psi_1(\mathrm{x}) + \sqrt{\frac{5}{7}}\Psi_2(\mathrm{x})$, many students incorrectly answered "2/7" in the pretest. In the post-test, however, some demonstrated a shift toward correct reasoning by stating that the probability of measuring $E_1$ would be unknown or very small because "any energy can be measured." These responses reflect a better understanding of the fact that a position eigenstate is a linear superposition of many energy eigenstates. Similarly, for question 4e, which asks whether the probability of measuring $E_1$ after a position measurement depends on time, some students stated in the pretest that it would not depend on time because the position measurement had collapsed the state. Although some of these students still did not arrive at the correct answer in the post-test, incorrectly stating that the probability of measuring $E_1$ would depend on time because "the system will evolve with time," their reasoning indicated a shift in understanding toward recognizing that position eigenstates evolve with time in a nontrivial manner. These examples suggest that even when students did not reach the correct answers in the post-test, we find some evidence at least for some students that their understanding may have moved closer to what would make it easier to help them learn the correct concepts.

While the findings of this study are promising, there are several limitations to consider. One limitation is that the post-test was administered shortly after the completion of the MQS sessions, which might raise concerns about whether the improved post-test performance is due to short-term memory effects rather than a more enduring understanding. Although this issue should be investigated in later studies, our experience suggests that this is unlikely to be a major factor, as students typically struggle with similar questions even when they have previously encountered the correct answers. Also, the post-test, which was implemented in the next class period after the MQS, had an open-ended format for the questions, which required students to explain their reasoning. This would also make it more challenging for students to rely on memorization. Moreover, we did observe a shift in student understanding toward correct reasoning by analyzing their responses to the test as discussed in the result section. Although this study did not include a retention test, our previous research with retention tests conducted at the end of the course has shown that students retain their learning gains over a longer period after engaging with MQS sessions [65,66]. Another limitation is the quasi-experimental design of the study. We did not include a control group that received only traditional instruction without the MQS intervention. However, it is important to note that the only assignment during the week of MQS implementation was traditional textbook homework. While we cannot entirely dismiss the potential influence of the homework, our previous research on tutorials and clicker questions in quantum mechanics consistently demonstrates that when controlling for factors like textbook study or homework [47], groups engaging with clicker questions still show greater learning gains. Given this, we believe the positive outcomes observed in this study are likely attributable to the MQS intervention.

## Acknowledgments

This research was carried out in accordance with the principles outlined in the University of Pittsburgh Institutional Review Board (IRB) ethical policy. We thank the NSF for Grant No. PHY-2309260. We thank all students whose data were analyzed and Dr. Robert P. Devaty for his constructive feedback on the manuscript. We also thank many members of the physics education group at Pitt for their help.

## Appendix A Quantum Measurement MQS

The multiple-choice questions in the sequence and notes to the instructors are reproduced below. The correct answers are in boldface.

Notes to Instructor

- In general, an operator acting on a state can be represented in Dirac notation as $\hat{Q}|\Psi\rangle$, where $\hat{Q}$ stands for a quantum operator, and $|\Psi\rangle$ stands for a quantum state.
- In this Multiple-choice Question Sequence (MQS), we write $\hat{Q}|\Psi\rangle$ in position representation as $\langle x|\hat{Q}|\Psi\rangle = \hat{Q}\left(x, -i\hbar\frac{\partial}{\partial x}\right)\Psi(\mathrm{x}) = \hat{Q}\Psi(\mathrm{x})$, where $\hat{Q}\left(x, -i\hbar\frac{\partial}{\partial x}\right)$ (shorten as $\hat{Q}$) is the operator $\hat{Q}$ in position representation, and $\Psi(\mathrm{x})$ is the wavefunction in position representation corresponding to state $|\Psi\rangle$.
- In order to go from the Dirac notation to position representation, one should take the scalar product with $\langle x|$

$$\hat{Q}|\Psi\rangle \longrightarrow \langle x|\hat{Q}|\Psi\rangle \stackrel{\text{def}}{=} \hat{Q}\left(x, -i\hbar\frac{\partial}{\partial x}\right)\Psi(x)$$

$$|\Psi\rangle \longrightarrow \langle x|\Psi\rangle \stackrel{\text{def}}{=} \Psi(x)$$

- In this Multiple-choice Question Sequence below, none of the operators explicitly depend on time.

**(MQS 1.1)**

Choose the following that is correct regarding the Hamiltonian operator $\hat{H}$ acting on a generic wavefunction $\Psi(x)$ which is not an eigenstate of $\hat{H}$.

A. $\hat{H}\Psi(x) = E\Psi(x)$
B. $\hat{H}\Psi(x) = E_n\psi_n(x)$
C. $\hat{H}\Psi(x) = E_n$
D. $\hat{H}\Psi(x) = \psi_n(x)$
***E. None of the above***

**Class discussion for MQS 1.1**

- Some students may incorrectly think that an operator acting on a state corresponds to a measurement of the corresponding observable and that this process of measuring the observable is given, e.g., by equations such as $\hat{H}\Psi(x) = E_n\psi_n(x)$ for measurement of energy.
- Some students may incorrectly think that whenever an operator $\hat{Q}$ corresponding to a physical observable Q acts on any generic state $\Psi(x)$, it will yield a corresponding eigenvalue and the same state back, i.e., $\hat{Q}\,\Psi(x) = \lambda\,\Psi(x)$ (e.g., $\hat{H}\Psi(x) = E\Psi(x)$)
- But only when $\Psi_q(x)$ is an eigenstate of $\hat{Q}$ , we obtain $\hat{Q}\,\Psi_q(x) = q\Psi_q(x)$, where $q$ is the corresponding eigenvalue.
- A generic wave function $\Psi(x)$ can be expanded in terms of a complete set of eigenstates of any Hermitian operator corresponding to an observable, e.g., if $\hat{H}\psi_n(x) = E_n\psi_n(x)$
    - $\Psi(x) = \sum_n C_n\,\psi_n(x)$, where $C_n = \langle\psi_n|\Psi\rangle$

Thus, $\hat{H}$ acting on a generic state $\Psi(x)$ can be represented by $\hat{H}\Psi(x) = \sum_n \hat{H}\, C_n\, \psi_n(x) = \sum_n E_n C_n\, \psi_n(x)$

- If the instructor has covered Dirac notation, they can also discuss:
  Because $\Psi(x) = \langle x|\Psi\rangle, |\Psi\rangle = \sum_n C_n|\psi_n\rangle$ , and $\hat{H}|\psi_n\rangle = E_n|\psi_n\rangle$

**(MQS 1.2)**
Choose the statement that is correct regarding the position operator $\hat{x}$ acting on a generic wavefunction $\Psi(x)$.

A. $\hat{x}\Psi(x) = x'$
B. $\hat{x}\Psi(x) = x'\delta(x - x')$
***C.*** $\boldsymbol{\hat{x}\Psi(x) = x\Psi(x)}$
D. $\hat{x}\Psi(x) = \delta(x - x')$
E. None of the above

**Class discussion for MQS 1.2**
This question is similar in spirit to the previous one about the Hamiltonian operator.
In the position representation, $\hat{x}\Psi(x) = x\Psi(x)$. This is because in position representation, $\hat{x} = x$. This does not mean that $\Psi(x)$ is an eigenstate of $\hat{x}$.

- If you have covered Dirac notation, you can also discuss:
  Because $\Psi(x) = \langle x|\Psi\rangle, |\Psi\rangle = \int_{-\infty}^{\infty} \Psi(x')\, |x'\rangle dx'$, $\hat{x}\, |x'\rangle = x'|x'\rangle$ and $\langle x'|\hat{x} = x'\langle x'|$
- $\hat{x}\delta(x - x') = \langle x|\hat{x}|x'\rangle = x'\langle x|x'\rangle = x'\delta(x - x')$ is an eigenvalue equation, in which position operator $\hat{x}$ is acting on a position eigenstate $\delta(x - x')$ (in the position representation) with eigenvalue $x'$. Here, $x'$ is a number, which is the eigenvalue corresponding to the position eigenstate $\delta(x - x')$.
  Please note that $\hat{x}\Psi(x) = x\Psi(x)$ is not an eigenvalue equation. In position representation, position operator $\hat{x}$ acting on any generic wave function $\Psi(x)$ or $\langle x|\Psi\rangle$ simply corresponds to multiplication by x.
- Or you can start with a generic expression of the position operator $\hat{x}$ acting on a state $|\Psi\rangle$, which is $\hat{x}|\Psi\rangle$. By multiplying it the bra $\langle x|$, we can write this expression in the position representation $\langle x|\hat{x}|\Psi\rangle$. Since $\langle x|\hat{x}|\Psi\rangle = x\langle x|\Psi\rangle = x\Psi(x)$, position operator $\hat{x}$ acting on any generic wave function $\Psi(x)$ or $\langle x|\Psi\rangle$ simply corresponds to multiplication by x, i.e., $\hat{x}\Psi(x) = x\Psi(x)$.

**Checkpoints**

- An operator (corresponding to an observable) acting on a state does NOT correspond to the measurement of the corresponding observable.
- What is the result of $\hat{x}$ acting on a position eigenstate?
- What is the result of $\hat{x}$ acting on a generic state?
- What is the result of $\hat{H}$ acting on an energy eigenstate?
- What is the result of $\hat{H}$ acting on a generic state?

**(MQS 2.1)**
Suppose at time $t = 0$, the initial wavefunction of a particle in a 1D infinite square well of width a (0<x<a) is $\Psi(x) = \frac{1}{\sqrt{2}}(\Psi_1(x) + \Psi_2(x))$, where $\Psi_1(x)$ and $\Psi_2(x)$ are the ground state and first excited state wavefunctions. Choose all of the following statements that are correct for measurements on the system in this state at $t = 0$.

1. A measurement of the energy can yield any energy $E_n$, where n=1,2,3…∞.
2. A measurement of the energy will yield $(E_1 + E_2)/2$.
3. A measurement of the position in a narrow range $dx$ can yield many different values in this well (0<x<a).

*A.* 1 only B. 2 only ***C. 3 only***
D. 2 and 3 only E. None of the above

**(MQS 2.2)**

Q is a generic observable (with corresponding Hermitian operator $\hat{Q}$ which has eigenstates $\varphi_q(x)$ and continuous eigenvalues q and eigenvalue equation $\hat{Q}\varphi_q(x) = q\varphi_q(x)$). Choose all of the following statements that are correct about a measurement of the observable $Q$ on a generic state $\Psi(x)$ (which is not an eigenstate of the operator $\hat{Q}$).

1. The measurement of the observable $Q$ will collapse the wavefunction into one of the eigenstates $\varphi_q(x)$ of operator $\hat{Q}$.
2. A measurement of an observable $Q$ must return one of the eigenvalues $q$ of the operator $\hat{Q}$.
3. The operator $\hat{Q}$ acting on state $\Psi(x)$ is equivalent to the measurement of the observable $Q$ . The measurement process is given by $\hat{Q}\Psi(x) = q\varphi_q(x)$.

A) 1 only ***B) 1 and 2 only*** C) 1 and 3 only
D) 2 and 3 only E) All of the above

**(MQS 2.3)**

At time $t = 0$, the initial wavefunction of a particle is $\Psi(x, 0) = (\frac{1}{5} - \frac{4}{5}i)\Psi_1(x) + \frac{\sqrt{8}}{5}\Psi_2(x)$, where $\Psi_1(x)$ and $\Psi_2(x)$ are the ground state and first excited state wavefunctions. Choose all of the following statements that are correct for measurements on the system in this state at $t = 0$.

1. If energy is measured, the probability of obtaining $E_1$ is $\frac{1}{25}$ and $E_2$ is $\frac{8}{25}$.
2. If position is measured, the probability density for measuring $x_0$ is $(\frac{17}{25}|\Psi_1(x_0)|^2 + \frac{8}{25}|\Psi_2(x_0)|^2)$
3. If a generic observable D (with corresponding Hermitian operator $\hat{D}$ which has eigenstates $|\phi_i\rangle$ and discrete eigenvalues $d_i$ and eigenvalue equation $\hat{D}|\phi_i\rangle = d_i|\phi_i\rangle$, where $i$ =1,2,3…) is measured, the probability of obtaining $d_i$ is $|\langle\phi_i|\Psi\rangle|^2$, where $|\Psi\rangle$ represents the quantum state corresponding to $\Psi(x)$.
4. If a generic observable Q (with corresponding Hermitian operator $\hat{Q}$ which has eigenstates $|q\rangle$ and continuous eigenvalues $q$ and eigenvalue equation $\hat{Q}|q\rangle = q|q\rangle$) is measured, the probability density for measuring $q$ is $|\langle q|\Psi\rangle|^2$, where $|\Psi\rangle$ represents the quantum state corresponding to $\Psi(x)$.

***A.*** 1 only B. 2 only ***C. 3 and 4 only***
D. 2 and 3 only E. None of the above

**Class discussion for MQS 2.3**

If you haven't discussed Dirac notation with the students, please feel free to replace the choice 3 and choice 4 with:

3. If a generic observable D (with corresponding Hermitian operator $\hat{D}$ which has eigenstates $\phi_i(x)$ and discrete eigenvalues $d_i$ and eigenvalue equation $\hat{D}\phi_i(x) = d_i\phi_i(x)$, where $i$ =1,2,3,4…) is measured, the probability of obtaining $d_i$ is $\left|\int_{-\infty}^{\infty} \phi_i^*(x)\Psi(x)dx\right|^2$.
4. If a generic observable Q (with corresponding Hermitian operator $\hat{Q}$ which has eigenstates $\varphi_q(x)$ and continuous eigenvalues $q$ and eigenvalue equation $\hat{Q}\varphi_q(x) = q\varphi_q(x)$) is measured, the probability density for measuring $q$ is $\left|\int_{-\infty}^{\infty} \varphi_q^*(x)\Psi(x)dx\right|^2$.

**Note:**

- In $\hat{D}|\phi_i\rangle = d_i|\phi_i\rangle$, operator $\hat{D}$ is an operator corresponding to observable D, and $|\phi_i\rangle$ stands for an eigenstate of $\hat{D}$ with eigenvalue $d_i$
- $\hat{D}\phi_i(x) = d_i\phi_i(x)$ is the same eigenvalue equation written in position representation, where $\hat{D}$ is $\hat{D}\left(x, -i\hbar\frac{\partial}{\partial x}\right)$, and $\phi_i(x)$ is an eigenfunction of $\hat{D}$ with eigenvalue $d_i$ in the position representation.

**Class discussion for MQS 2.1-2.3**

Consider a generic wavefunction $\Psi(x)$ corresponding to a generic state $|\Psi\rangle$.

- If you measure energy, the probability of obtaining $E_m$ is
  $P(E_m) = |\langle\Psi_m|\Psi\rangle|^2 = \left|\int_{-\infty}^{\infty}\langle\Psi_m|x\rangle\langle x|\Psi\rangle dx\right|^2 = \left|\int_{-\infty}^{\infty} \Psi_m^*(x)\Psi(x)dx\right|^2$ (where $|\Psi_m\rangle$ is an energy eigenstate satisfying $\hat{H}|\Psi_m\rangle = E_m|\Psi_m\rangle$)
- If you measure position, the probability density for measuring $x_0$ is
  $\rho(x_0) = |\langle x_0|\Psi\rangle|^2 = \left|\int_{-\infty}^{\infty}\langle x_0|x\rangle\langle x|\Psi\rangle dx\right|^2 = \left|\int_{-\infty}^{\infty} \delta(x_0 - x)\Psi(x)dx\right|^2 = |\Psi(x_0)|^2$ (where $|x_0\rangle$ is a position eigenstate satisfying $\hat{x}|x_0\rangle = x_0|x_0\rangle$)
- If you measure the observable D whose corresponding Hermitian operator $\hat{D}$ has a discrete eigenvalue spectrum, the probability of obtaining $d_i$ is
  $P(d_i) = |\langle\phi_i|\Psi\rangle|^2 = \left|\int_{-\infty}^{\infty}\langle\phi_i|x\rangle\langle x|\Psi\rangle dx\right|^2 = \left|\int_{-\infty}^{\infty} \phi_i^*(x)\Psi(x)dx\right|^2$ (where $|\phi_i\rangle$ is an eigenstate of $\hat{D}$ satifying $\hat{D}|\phi_i\rangle = d_i|\phi_i\rangle$)
- If you measure the observable Q whose corresponding operator $\hat{Q}$ has a continuous eigenvalue spectrum, the probability density for measuring $q$ is
  $\rho(q) = |\langle q|\Psi\rangle|^2 = \left|\int_{-\infty}^{\infty}\langle q|x\rangle\langle x|\Psi\rangle dx\right|^2 = \left|\int_{-\infty}^{\infty} \varphi_q^*(x)\Psi(x)dx\right|^2$ (where $|q\rangle$ is an eigenstate of $\hat{Q}$ satifying $\hat{Q}|q\rangle = q|q\rangle$)

**Note:**

1. Emphasize to students that when considering the probability of measuring an observable in a generic state, they should be thinking about the measurement basis, which is the basis consisting of the eigenstates of the operator corresponding to the observable being measured. Then projecting the generic state onto an eigenstate of the operator corresponding to the observable (in the measurement basis), the absolute square of the projection will give the probability (discrete eigenvalues) or probability density (continuous eigenvalues) of obtaining the corresponding eigenvalue.

2. The projections are given by the inner product of the generic state $\Psi(x)$ and the eigenstates of the operator corresponding to the observable. For example, for energy measurement, the projection of state $\Psi(x)$ onto energy eigenstate $\Psi_m(x)$ is $\int_{-\infty}^{\infty} \Psi_m^*(x)\Psi(x)dx$. If you have discussed Dirac notation, this projection can be simply represented by $\langle\Psi_m|\Psi\rangle$.
3. We can calculate the probability/probability density by projecting the generic state onto the eigenstates as discussed above. However, if we can expand the generic state in terms of a complete set of eigenstates of the operator corresponding to the observable we measure, the probability/probability density of measuring a particular value of the observable is the absolute square of the corresponding expansion coefficient.

   - For example, we can expand a generic state $\Psi(x)$ in terms of energy eigenstates, $\Psi(x) = \sum_n C_n \Psi_n(x)$, then $P(E_m) = |\langle\Psi_m|\Psi\rangle|^2 = \left|\int_{-\infty}^{\infty}\langle\Psi_m|x\rangle\langle x|\Psi\rangle dx\right|^2 = \left|\int_{-\infty}^{\infty} \Psi_m^*(x)\Psi(x)dx\right|^2 = \left|\sum_n C_n \int_{-\infty}^{\infty} \Psi_m^*(x)\Psi_n(x)dx\right|^2 = |\sum_n C_n\, \delta_{mn}|^2 = |C_m|^2$
   - If you have done Dirac notation with students, you can also calculate the probability in the following way:
     $P(E_m) = |\langle\Psi_m|\Psi\rangle|^2 = |\sum_n C_n\langle\Psi_m|\Psi_n\rangle|^2 = |\sum_n C_n\, \delta_{mn}|^2 = |C_m|^2$
4. The probability density for measuring position $x_0$ is $\rho(x_0) = |\langle x_0|\Psi\rangle|^2 = |\Psi(x_0)|^2$

**Checkpoints**

- When a measurement is made, what measured values can you obtain if the state of the system is not an eigenstate of the operator corresponding to an observable (or when it IS an eigenstate of the operator corresponding to the observable)?
- When a measurement is made, what happens to the wave function instantaneously after the measurement if the state is not an eigenstate of the operator corresponding to the observable measured?
- Does a Hermitian operator corresponding to an observable acting on the quantum state correspond to a measurement of the observable? No!
- How does one calculate probability (discrete eigenvalue spectrum) or probability density (continuous eigenvalue spectrum) of measuring an observable?

**(MQS 3.1)**

Choose all of the following statements that are correct.

1. The stationary states refer to the eigenstates of any operator corresponding to a physical observable.
2. Any wavefunction for a system can be expressed as a linear superposition of the energy eigenstates.
3. If a system is in an eigenstate of any operator that corresponds to a physical observable, it stays in that state unless an external perturbation is applied.

A. 1 only ***B. 2 only*** C. 3 only
D. 2 and 3 only E. None of the above

**(MQS 3.2)**

Suppose at time $t = 0$, a particle in a 1D infinite square well has the initial wavefunction $\Psi(x,0) = \frac{1}{\sqrt{2}}(\Psi_1(x) + \Psi_2(x))$, where $\Psi_1(x)$ and $\Psi_2(x)$ are the ground state and first excited state wavefunctions. Choose all of the following expressions that can correctly represent the state $\Psi(x,t)$ of the particle after time *t*

1. $\Psi(x,t) = \frac{1}{\sqrt{2}} e^{\frac{-i(E_1+E_2)\,t}{2\hbar}} (\Psi_1(x) + \Psi_2(x))$
2. $\Psi(x,t) = \frac{1}{\sqrt{2}} (e^{\frac{-iE_1 t}{\hbar}} \Psi_1(x) + e^{\frac{-iE_2 t}{\hbar}} \Psi_2(x))$
3. $\Psi(x,t) = \frac{1}{\sqrt{2}} e^{\frac{-i\hat{H}t}{\hbar}} (\Psi_1(x) + \Psi_2(x))$

A. 1 only B. 2 only C. 1 and 3 only
***D. 2 and 3 only*** E. None of the above

**Class discussion for MQS 3.2**

- By solving the Time-dependent Schrödinger equation $i\hbar \frac{\partial}{\partial t} \Psi(x,t) = \hat{H}\, \Psi(x,t)$ , the time dependance of a generic state is given by $\Psi(x,t) = e^{\frac{-i\hat{H}t}{\hbar}} \Psi(x,0)$, where $e^{\frac{-i\hat{H}t}{\hbar}}$ is the time-evolution operator.
- Any wavefunction for a system can be expressed as a linear superposition of the energy eigenstates:

$\Psi(x,0) = \sum_n C_n\, \psi_n(x)$, where $\psi_n(x)$ is an energy eigenstate with eigenvalue $E_n$. Thus,

$$\Psi(x,t) = e^{\frac{-i\hat{H}t}{\hbar}} \Psi(x,0) = \sum_n C_n e^{\frac{-i\hat{H}t}{\hbar}} \psi_n(x)$$

Let's look at one term in this expansion $-\ C_n e^{\frac{-i\hat{H}t}{\hbar}} \psi_n(x)$. We can expand the exponential function $e^{\frac{-i\hat{H}t}{\hbar}}$ as follow: $e^{\frac{-i\hat{H}t}{\hbar}} = 1 + \frac{-i\hat{H}t}{\hbar} + \frac{1}{2}\left(\frac{-i\hat{H}t}{\hbar}\right)^2 + \frac{1}{6}\left(\frac{-i\hat{H}t}{\hbar}\right)^3 + \cdots$

Thus, $e^{\frac{-i\hat{H}t}{\hbar}} \psi_n(x) = \psi_n(x) + \frac{-i\hat{H}t}{\hbar} \psi_n(x) + \frac{1}{2}\left(\frac{-i\hat{H}t}{\hbar}\right)^2 \psi_n(x) + \frac{1}{6}\left(\frac{-i\hat{H}t}{\hbar}\right)^3 \psi_n(x) + \cdots$

Because $\hat{H}\psi_n(x) = E_n \psi_n(x)$, we have $\frac{-i\hat{H}t}{\hbar} \psi_n(x) = \frac{-iE_n t}{\hbar} \psi_n(x)$, $\left(\frac{-i\hat{H}t}{\hbar}\right)^2 \psi_n(x) = \left(\frac{-iE_n t}{\hbar}\right)^2 \psi_n(x)$, $\left(\frac{-i\hat{H}t}{\hbar}\right)^3 \psi_n(x) = \left(\frac{-iE_n t}{\hbar}\right)^3 \psi_n(x), \ldots$

Therefore, $e^{\frac{-i\hat{H}t}{\hbar}} \psi_n(x) = \psi_n(x) + \frac{-iE_n t}{\hbar} \psi_n(x) + \frac{1}{2}\left(\frac{-iE_n t}{\hbar}\right)^2 \psi_n(x) + \frac{1}{6}\left(\frac{-iE_n t}{\hbar}\right)^3 \psi_n(x) + \cdots = e^{\frac{-iE_n t}{\hbar}} \psi_n(x)$

Thus, $\Psi(x,t) = e^{\frac{-i\hat{H}t}{\hbar}} \Psi(x,0) = \sum_n C_n e^{\frac{-i\hat{H}t}{\hbar}} \psi_n(x) = \sum_n C_n e^{\frac{-iE_n t}{\hbar}} \psi_n(x)$

Therefore, the initial state $\Psi(x,0)$ evolves in time such that each term in the expansion of the state in terms of the energy eigenstates is multiplied by its corresponding time dependent factor $e^{\frac{-iE_n t}{\hbar}}$. Note that in general, the time dependent factor is different for each term because the energy corresponding to each term is generally different.

- As a special case, if $\Psi(x,0)$ itself is an energy eigenstate with eigenvalue $E_n$, then $\Psi(x,0) = \psi_n(x)$ and $\Psi(x,t) = e^{\frac{-i\hat{H}t}{\hbar}} \Psi(x,0) = e^{\frac{-i\hat{H}t}{\hbar}} \psi_n(x) = e^{\frac{-iE_n t}{\hbar}} \psi_n(x)$. In this case, the time-evolution of the state is trivial because the state is multiplied by an overall phase factor.

**(MQS 3.3)**

Suppose at time $t = 0$, the initial wavefunction of a particle in a 1D infinite square well is $\Psi(x) = \frac{1}{\sqrt{2}}(\Psi_1(x) + \Psi_2(x))$, where $\Psi_1(x)$ and $\Psi_2(x)$ are the ground state and first excited state wavefunctions. Choose all of the following statements that are correct for measurements on the system in this state at $t = 0$.

1. If **energy** is measured, the wavefunction will become either $\Psi_1(x)$ or $\Psi_2(x)$ immediately after the energy measurement but go back to $\Psi(x) = \frac{1}{\sqrt{2}}(\Psi_1(x) + \Psi_2(x))$ a long time after the measurement.
2. If **position** is measured, the wavefunction $\Psi(x) = \frac{1}{\sqrt{2}}(\Psi_1(x) + \Psi_2(x))$ will become a delta function immediately after the position measurement, but go back to $\Psi(x) = \frac{1}{\sqrt{2}}(\Psi_1(x) + \Psi_2(x))$ a long time after the measurement.
3. If **position** is measured, the wavefunction $\Psi(x) = \frac{1}{\sqrt{2}}(\Psi_1(x) + \Psi_2(x))$ will become a delta function immediately after the position measurement, and the wavefunction will remain the delta function a long time after the measurement.

A. 2 only B. 3 only C. 1 and 3 only
D. 1 and 2 only ***E. None of the above***

**Class discussion for MQS 3.3**

It should be emphasized to students that whenever talking about time evolution, we should expand the wave function in terms of energy eigenstates, and then multiply each term in the expansion by the corresponding time-dependent phase factor.

- If we measure energy at $t = 0$ and obtain $E_n$, the wavefunction will instantaneously collapse to $\psi_n(x)$. After time $t_1$, this wave function becomes $e^{\frac{-iE_n t_1}{\hbar}}\psi_n(x)$. Even though there is a time dependent phase factor, $e^{\frac{-iE_n t_1}{\hbar}}\psi_n(x)$ is still an energy eigenstate with eigenvalue $E_n$. Thus, the spatial part of the wavefunction remains $\psi_n(x)$ after the energy measurement.

$$t = 0 \qquad\qquad\qquad t = t_1$$

$$\Psi(x,0) = \psi_n(x) \;-\!-\!-\!-\!-\!-\!-\!-\!\rightarrow\; \Psi(x,t_1) = e^{\frac{-iE_n t_1}{\hbar}}\psi_n(x)$$

**Note:**

- Multiplying a state with an overall phase factor doesn't change the state. Therefore, $e^{\frac{-iE_n t_1}{\hbar}}\psi_n(x)$ shows the time dependence of the stationary state with energy $E_n$.
- For example, if we measure energy at $t = 0$ and obtain $E_3$, the wave function will collapse to $\psi_3(x)$ right after an energy measurement. The three time-lapsed pictures from a simulation below (Figure 1) show the time evolution of the absolute value of $\psi_3(x)$ in a 1D infinite square well with boundary ($0<x<a$). (a) shows the absolute value of the wavefunction right after the energy measurement. (b) and (c) show the absolute value of the wave function at $t = 0.3$ units and $t = 14.3$ units. As shown in the simulation, the absolute value of the wavefunction does not change with time. Thus, the wavefunction will remain $\psi_3(x)$ after the energy measurement up to a trivial overall time-dependent phase factor.

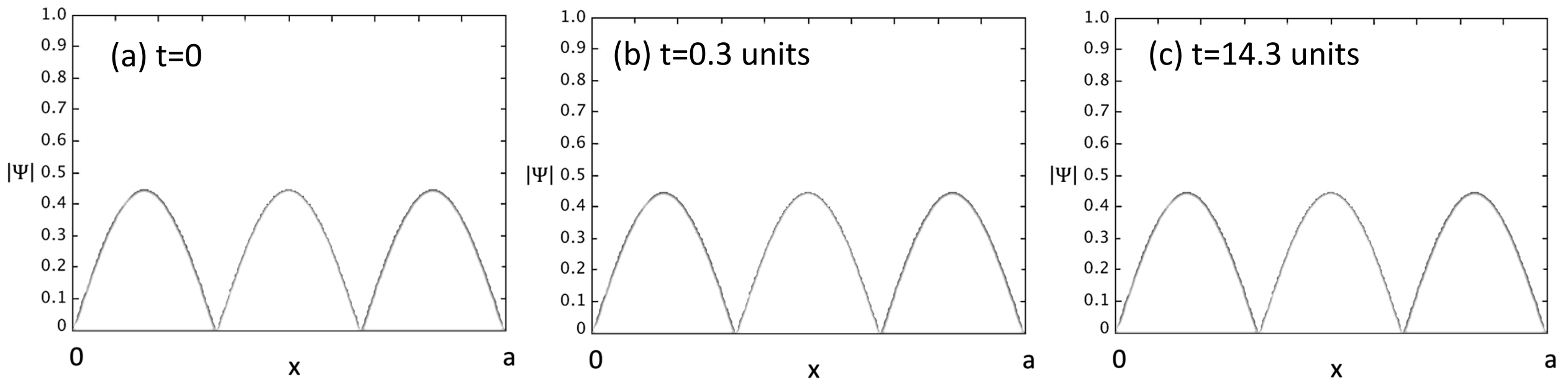

Figure 1. Time-lapsed images from a simulation depicting the time evolution of the absolute value of $\psi_3(x)$ in a 1D infinite square well with boundary (0<x<a). (a) shows the absolute value of the wavefunction right after the energy measurement. (b) and (c) show the absolute value of the wave function at *t* = 0.3 units and *t* = 14.3 units.

- If we measure position at *t* = 0 and obtain $x_0$, the wavefunction will instantaneously collapse to $\delta(x-x_0)$, which can be expanded as $\sum_n C_n\,\psi_n(x)$. After time $t_1$, the wave function becomes $\sum_n C_n\, e^{\frac{-iE_nt_1}{\hbar}}\psi_n(x)$. Thus, the wavefunction will neither remain $\delta(x-x_0)$ nor go back to $\frac{1}{\sqrt{2}}\big(\Psi_1(x)+\Psi_2(x)\big)$.

$$\begin{array}{ccc} t=0 & & t=t_1 \\ \Psi(x,0)=\delta(x-x_0)=\sum_n C_n\,\psi_n(x) & --------\rightarrow & \Psi(x,t_1)=\sum_n C_n\, e^{\frac{-iE_nt_1}{\hbar}}\psi_n(x) \end{array}$$

**Note:**

- The expansion coefficient in this case is $C_n=\int_{-\infty}^{\infty}\Psi(x,0)\psi_n^*(x)\mathrm{dx}=\int_{-\infty}^{\infty}\delta(x-x_0)\psi_n^*(x)\mathrm{dx}=\psi_n^*(x_0)$.
- In general, $\sum_n C_n\, e^{\frac{-iE_nt_1}{\hbar}}\psi_n(x)$ is not equal to $\delta(x-x_0)=\sum_n C_n\,\psi_n(x)$ , because $e^{\frac{-iE_nt_1}{\hbar}}$ are generally different for each $E_n$.
- The three time-lapsed pictures from a simulation below (Figure 2) show the time evolution of $\delta(x-x_0)$ in a 1D infinite square well with boundary (0<x<a). (a) shows the absolute value of the wave function which is very localized (the actual wavefunction right after a position measurement is closer to a delta function which will be highly localized but the simulation is unable to show that level of a localized function so students would have to imagine a very peaked function). (b), (c) show the absolute value of the wave function at *t* = 0.6 units and *t* = 1.6 units. As shown in the simulation, the wave function does not remain localized (similarly, a delta function $\delta(x-x_0)$ after the position measurement will not remain localized at future times).
- Since delta function $\delta(x-x_0)$ contains nonzero coefficients $C_n$ for higher energy eigenfunction $\psi_n(x)(n>2)$, the probability of measuring these higher energies $\left|C_n e^{\frac{-iE_nt_1}{\hbar}}\right|^2$ would not be zero at future times. Therefore, the system will not return to the initial state $\frac{1}{\sqrt{2}}\big(\Psi_1(x)+\Psi_2(x)\big)$ after the position measurement, no matter how long we wait.

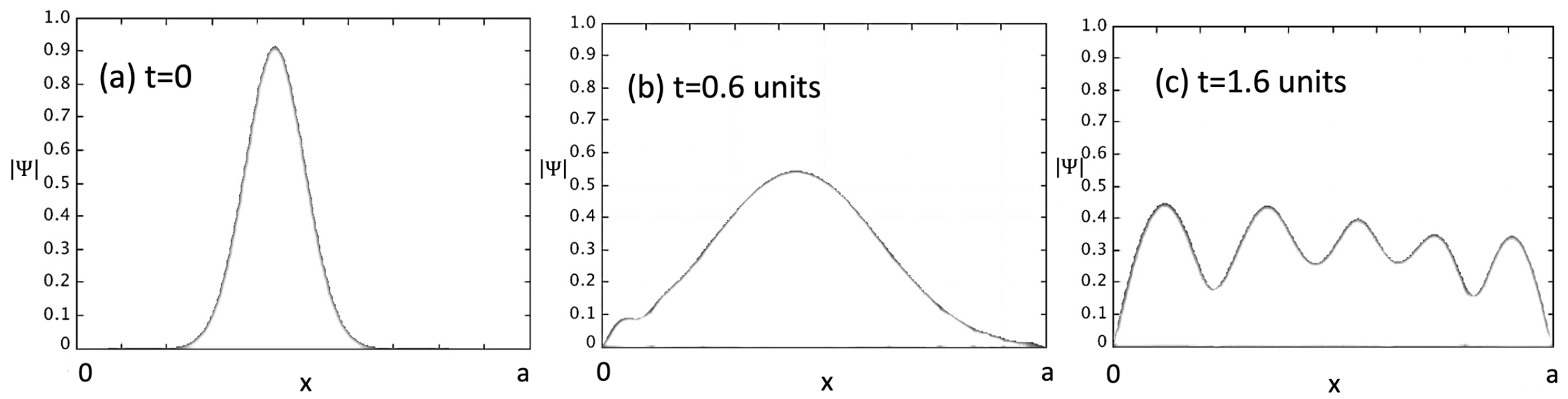

Figure 2. Time-lapsed images from a simulation below show the time evolution of a very peaked function (since the simulation is unable to show a delta function) in a 1D infinite square well with boundary (0<x<a). (a) shows the absolute value of the wave function which is very localized. (b), (c) show the absolute value of the wave function at *t* = 0.6 units and *t* = 1.6 units.

**(MQS 3.4)**

Suppose at time $t = 0$, a particle in a 1D infinite square well has the initial wavefunction $\Psi(x) = \frac{1}{\sqrt{2}}(\Psi_1(x) + \Psi_2(x))$, where $\Psi_1(x)$ and $\Psi_2(x)$ are the ground state and first excited state wavefunctions (see the left graph in Fig. 3 below). You measure the position of the particle and obtain $x_0$. Choose all of the following statements that are correct.

1. The wavefunction will instantaneously collapse to a delta function at $x = x_0$ (the middle graph in Fig.3 is a schematic) when the position measurement is performed.
2. The wavefunction will remain a delta function at $x_0$ (one example is the middle graph in Fig. 3) a long time after the position measurement.
3. The wavefunction must evolve from a delta function at $x_0$ (one example is the right graph in Fig. 3) so that it has different shapes a long time after the position measurement.

A. 1 only B. 1 and 2 only ***C. 1 and 3 only***
D. 2 and 3 only E. all of the above.

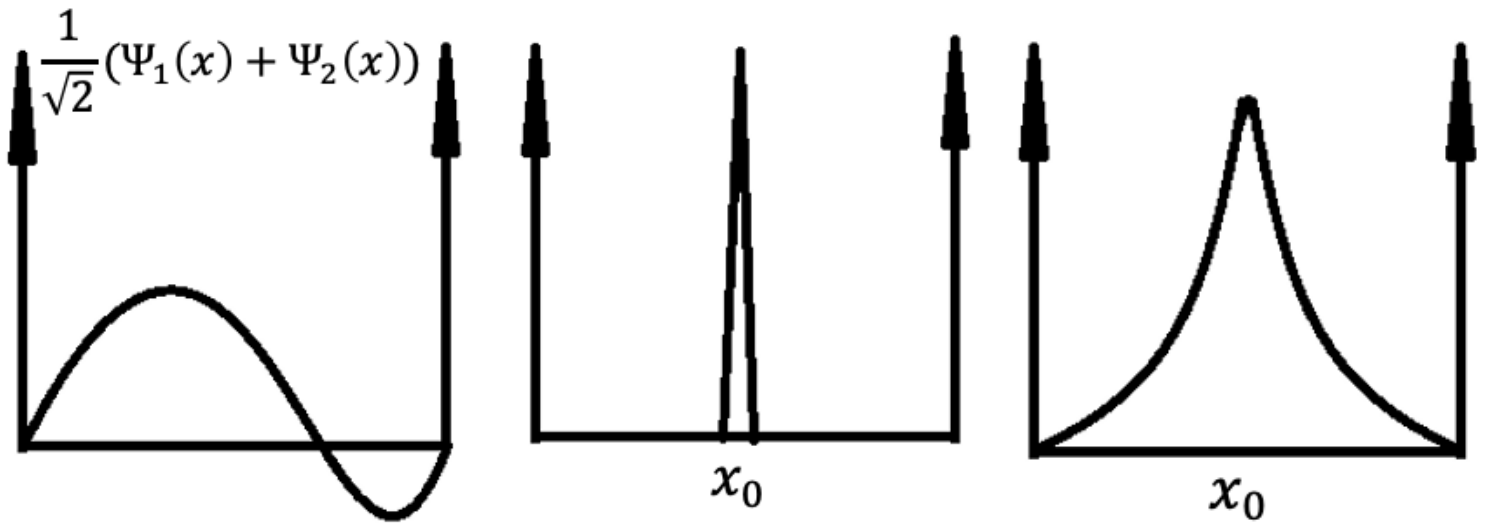

Figure 3. Figure for MQS 3.4.

**(MQS 3.5)**

Suppose at time $t = 0$, a particle in a 1D infinite square well (0 < x < a) has the initial wavefunction $\Psi(x) = \frac{1}{\sqrt{2}}(\Psi_1(x) + \Psi_2(x))$, where $\Psi_1(x)$ and $\Psi_2(x)$ are the ground state and first excited state wavefunctions. You measure the energy of the particle and obtain $E_2$. Choose all of the following statements that are correct.

1. The wavefunction will instantaneously collapse to a stationary state when the energy measurement yielding $E_2$ is performed (the middle graph in Fig. 4).
2. The wavefunction will remain the first excited state a long time after the energy measurement.
3. The wavefunction must return to its initial state before the measurement a long time after the energy measurement (the right graph in Fig. 4).

A. 1 only ***B. 1 and 2 only*** C. 3 only
D 1 and 3 only E. None of the above.

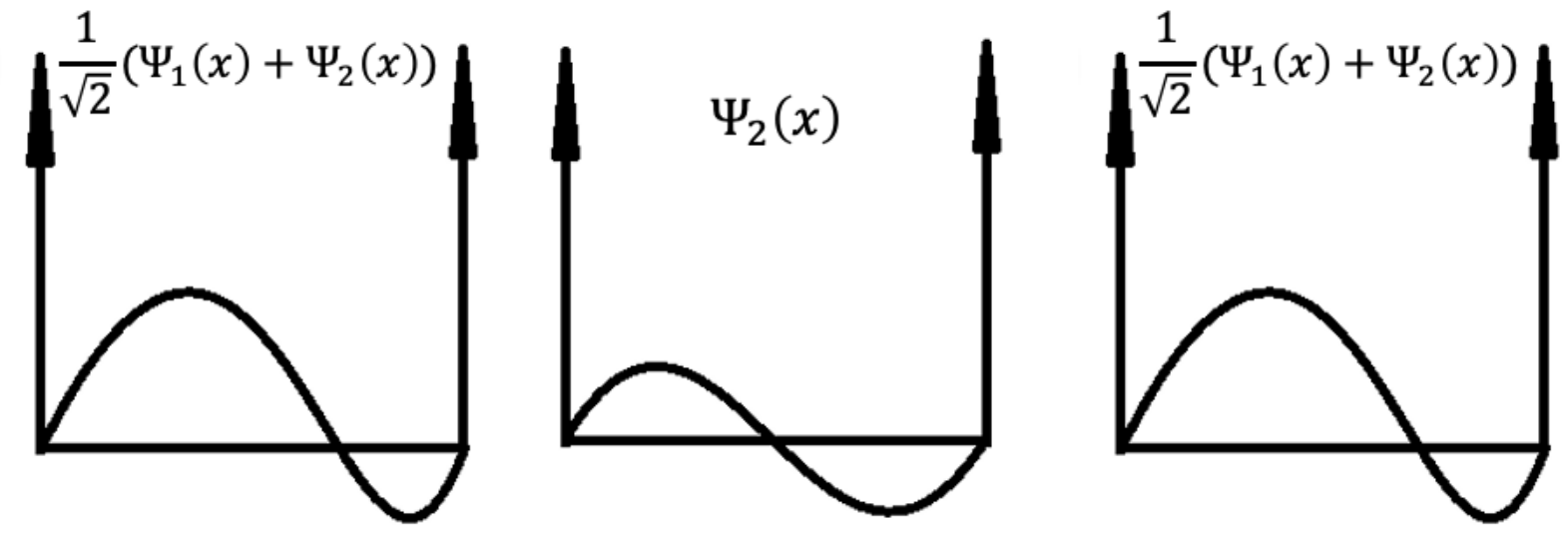

Figure 4. Figure for MQS 3.5.

**(MQS 3.6)**

At time $t = 0$, the initial wavefunction of a particle is $\Psi(\mathrm{x},\ 0) = \frac{1}{\sqrt{2}}(\Psi_1(\mathrm{x}) + \Psi_2(\mathrm{x}))$, where $\Psi_1(\mathrm{x})$ and $\Psi_2(\mathrm{x})$ are the ground and first-excited energy eigenstates. Choose all of the following statements that are correct about a measurement performed on this system at $t = t_1$.

1. If we measure the position of the particle, the probability density for measuring the position of the particle will depend on the time $t_1$.
2. If we measure the energy of the particle, the probability of obtaining the energy $E_1$ or $E_2$ will depend on the time $t_1$.
3. If we measure an observable Q (with corresponding Hermitian operator $\widehat{\mathrm{Q}}$ which has eigenstates $\varphi_\mathrm{q}(\mathrm{x})$ and continuous eigenvalues q and eigenvalue equation $\widehat{\mathrm{Q}}\varphi_\mathrm{q}(\mathrm{x}) = \mathrm{q}\varphi_\mathrm{q}(\mathrm{x})$), the probability density for measuring *q* will depend on the time $t_1$ (Note: $\widehat{\mathrm{Q}}$ does NOT commute with $\widehat{\mathrm{H}}$).

A. 1 only B. 2 only C. 3 only
***D. 1 and 3 only*** E. all of the above

**Class discussion for MQS 3.6**

Emphasize to students that when talking about time evolution, since the Hamiltonian of the system governs the time-evolution, we should expand the wave function in terms of energy eigenstates, and multiply each term by the corresponding time-dependent phase factor.

Initial state After waiting for $t$

$$\Psi(\mathrm{x},\ 0) = \sum_\mathrm{n} \mathrm{C_n}\, \psi_\mathrm{n}(\mathrm{x}) \text{ - - - - - - - - }\rightarrow \Psi(\mathrm{x},\mathrm{t}) = \sum_\mathrm{n} \mathrm{C_n}\, e^{\frac{-\mathrm{i}\mathrm{E_n t}}{\hbar}} \psi_\mathrm{n}(\mathrm{x})$$

- **If we measure energy at *t* = $t_1$, the probability of obtaining $E_\mathrm{m}$ is**

$P(E_m) = |\langle\Psi_m|\Psi(t_1)\rangle|^2 = \left|\int_{-\infty}^{\infty}\langle\Psi_m|x\rangle\langle x|\Psi(t_1)\rangle dx\right|^2 = |\int_{-\infty}^{\infty}\Psi_m^*(x)\Psi(x,t_1)dx|^2 =$ $\left|\sum_n C_n e^{\frac{-iE_nt_1}{\hbar}}\int_{-\infty}^{\infty}\Psi_m^*(x)\Psi_n(x)dx\right|^2 = \left|\sum_n C_n e^{\frac{-iE_nt_1}{\hbar}}\delta_{mn}\right|^2 = \left|C_m e^{\frac{-iE_mt_1}{\hbar}}\right|^2 = |C_m|^2$, which is independent of $t_1$.

- **If we measure position at $t = t_1$, the probability density for measuring $x_0$ is** $\rho(x_0) = |\langle x_0|\Psi(t_1)\rangle|^2 = \left|\int_{-\infty}^{\infty}\langle x_0|x\rangle\langle x|\Psi(t_1)\rangle dx\right|^2 = \left|\int_{-\infty}^{\infty}\delta(x_0 - x)\Psi(x,t_1)dx\right|^2 =$ $|\Psi(x_0,t_1)|^2 = \left|\sum_n C_n e^{\frac{-iE_nt_1}{\hbar}}\psi_n(x_0)\right|^2$, which depends on $t_1$.
- **If we measure Q at $t = t_1$, the probability density for measuring $q$ is** $\rho(q) = |\langle q|\Psi(t_1)\rangle|^2 = \left|\int_{-\infty}^{\infty}\langle q|x\rangle\langle x|\Psi(t_1)\rangle dx\right|^2 = \left|\int_{-\infty}^{\infty}\varphi_q^*(x)\Psi(x,t_1)dx\right|^2 =$ $\left|\sum_n C_n e^{\frac{-iE_nt_1}{\hbar}}\int_{-\infty}^{\infty}\varphi_q^*(x)\,\psi_n(x)dx\right|^2$
  Because $\hat{Q}$ does NOT commute with $\hat{H}$, $\hat{Q}$ and $\hat{H}$ don't have a complete set of simultaneous eigenstates. Therefore, the time factor in $\left|\sum_n C_n e^{\frac{-iE_nt_1}{\hbar}}\int_{-\infty}^{\infty}\varphi_q^*(x)\,\psi_n(x)dx\right|^2$ will not cancel out. Thus, $\rho(q)$ depends on $t_1$.

**Note:**

1. For the position measurement, the probability density for measuring position $x_0$ is $\rho(x_0) = |\langle x_0|\Psi(t_1)\rangle|^2 = |\Psi(x_0,t_1)|^2 = \left|\sum_n C_n e^{\frac{-iE_nt_1}{\hbar}}\psi_n(x_0)\right|^2$, which depends on time.
2. If you have discussed Dirac notation with students, you can also discuss calculation of the probability of obtaining $E_m$ in the following way: $P(E_m) = |\langle\Psi_m|\Psi(t_1)\rangle|^2 =$ $\left|\sum_n C_n e^{\frac{-iE_nt_1}{\hbar}}\langle\Psi_m|\Psi_n\rangle\right|^2 = \left|\sum_n C_n e^{\frac{-iE_nt_1}{\hbar}}\delta_{mn}\right|^2 = \left|C_m e^{\frac{-iE_mt_1}{\hbar}}\right|^2 = |C_m|^2$, which is independent of $t_1$.

**Checkpoints:**

Consider a generic state $\Psi(x,\ 0)$ at $t = 0$.

- If energy is measured, what is the wave function instantaneously after and long time after the energy measurement?
- If position is measured, what is the wave function instantaneously after and long time after the position measurement?

**(MQS 4.1)**

The energy eigenvalues for a one-dimensional simple harmonic oscillator (SHO) are $E_n = \left(n + \frac{1}{2}\right)\hbar\omega$ $(n = 0,1,2,3,\ldots)$. The initial wavefunction of a particle in a SHO potential energy well is $\Psi(x,\ 0) = \frac{1}{\sqrt{2}}(\Psi_1(x) + \Psi_2(x))$. You first measure the energy of the particle at $t = 0$ and obtain $\frac{3}{2}\hbar\omega$. Then immediately following the energy measurement, you measure the position of the particle. What is the probability of finding the particle in the region between $x_0$ and $x_0 + dx$?

1. $\frac{1}{2}(|\Psi_1(x_0)|^2+|\Psi_2(x_0)|^2)dx$
2. $|\Psi_1(x_0)|^2 dx$
3. $|\hat{x}\Psi_1(x)|^2 dx$

A. 1 only ***B. 2 only*** C. 3 only
D. 2 and 3 only E. None of the above

**Note:**
Emphasize to students that even though MQS 4.1 is about a quantum harmonic oscillator, the expression of probability density for measuring $x_0$ is generalizable to other quantum systems with different Hamiltonians such as the 1D infinite square well.

**(MQS 4.2)**
At time $t = 0$, the initial state of a particle in a 1D infinite square well is $\Psi(x, 0) = \frac{1}{\sqrt{2}}(\Psi_1(x) + \Psi_2(x))$, where $\Psi_1(x)$ and $\Psi_2(x)$ are the ground and first-excited energy eigenstates. You first measure the energy of the particle at $t = 0$ and obtain $E_1$. Then you measure the position of the particle at a later time $t = t_1$ (not immediately after the measurement of energy). What is the probability of finding the particle in the region between $x_0$ and $x_0 + dx$?

1. $\left|e^{\frac{-iE_1t_1}{\hbar}}\Psi_1(x_0)\right|^2 dx$
2. $\left|e^{\frac{-iE_1t_1}{\hbar}}\Psi_1(x_0) + e^{\frac{-iE_2t_1}{\hbar}}\Psi_2(x_0)\right|^2 dx$
3. $\left|\sum_n c_n e^{\frac{-iE_nt_1}{\hbar}}\Psi_n(x_0)\right|^2 dx$, where $c_n = \langle\Psi_n|\Psi\rangle$, n = 1, 2, 3…. and ($c_n \neq 0$ for 1, 2, 3….)

***A. 1 only*** B. 2 only C. 3 only
D. 1 or 2 depending on how much time has elapsed between the measurements
E. None of the above

**Note:**

- If you haven't discussed Dirac notation with students, please feel free to replace choice 3 with:
  $\left|\sum_n c_n e^{\frac{-iE_nt_1}{\hbar}}\Psi_n(x_0)\right|^2 dx$, where $c_n = \int_{-\infty}^{\infty}\Psi_n^*(x)\Psi(x)dx$, n = 1, 2, 3… and $c_n \neq 0$ for 1, 2, 3….
- It is worth noting that $\left|e^{\frac{-iE_1t_1}{\hbar}}\Psi_1(x_0)\right|^2 dx = |\Psi_1(x_0)|^2 dx$, which means that if the system is in a stationary state, the probability density for measuring position does not change with time.

**(MQS 4.3)**
The initial state of a particle in a 1D infinite square well (0 < x < a) is $\Psi(x, 0) = \frac{1}{\sqrt{2}}(\Psi_1(x) + \Psi_2(x))$. You first measure the position of the particle and obtain $x_0$. Then immediately following the position measurement, you measure the position of the particle again. Choose all of the following that are correct:

1. The second measurement must also yield $x_0$.
2. The second measurement could yield any of the infinitely many position eigenvalues possible for the system (0 < x < a).
3. The wavefunction immediately after the second measurement is still the position eigenstate corresponding to eigenvalue $x_0$.

A. 1 only B. 2 only C. 3 only
***D. 1 and 3 only*** E. 2 and 3 only

**(MQS 4.4)**
The initial state of a particle in a 1D infinite square well (0 < x < a) is $\Psi(x,\ 0) = \frac{1}{\sqrt{2}}(\Psi_1(x) + \Psi_2(x))$ when you measure the position of the particle and obtain $x_0$. Then some time *t* later following the position measurement, you measure the position of the particle again. Choose all of the following that are correct:

1. The second measurement must also yield $x_0$.
2. The second measurement could yield any of the infinitely many position eigenvalues possible for the system 0 < x < a.
3. The probability density for the second position measurement will depend on how much time elapses between the two measurements.

A. 1 only B. 2 only C. 3 only
D. 1 and 3 only ***E. 2 and 3 only***

**(MQS 4.5)**
The initial state of a particle in a 1D infinite square well (0 < x < a) is $\Psi(x,\ 0) = \frac{1}{\sqrt{2}}(\Psi_1(x) + \Psi_2(x))$ when you measure the position of the particle and obtain $x_0$. Then some time *t* later following the position measurement, you measure the energy of the particle. Choose all of the following that are correct:

1. The energy measurement can yield any energy $E_n$, where $n = 1,2,3 \ldots \infty$.
2. The wavefunction will become either $\Psi_1(x)$ or $\Psi_2(x)$ immediately after the energy measurement and the system will remain in that state at future times.
3. The probability of obtaining the energy $E_1$ will depend on how much time elapses between the two measurements.

***A. 1 only*** B. 2 only C. 3 only
D. 1 and 3 only E. 2 and 3 only

**Class discussion for MQS 4.1-4.5**
Emphasize to students that when talking about time evolution, we should expand the wave function in terms of energy eigenstates, and multiply each term by the corresponding time-dependent phase factor.

- **If we measure energy at *t* = 0 and obtain $E_n$**, the wavefunction will instantaneously collapse to $\psi_n(x)$. After time $t_1$, the wave function becomes $e^{\frac{-iE_n t_1}{\hbar}}\psi_n(x)$.

  $t = 0$ $t = t_1$

$$\Psi(x,0) = \psi_n(x) \text{ - - - - - - - -}\rightarrow \Psi(x,t_1) = e^{\frac{-iE_n t_1}{\hbar}}\psi_n(x)$$

- **If we measure energy at $t = t_1$**, the probability of obtaining $E_m$ is

  $P(E_m) = |\int_{-\infty}^{\infty} \Psi_m^*(x)\Psi(x,t_1)dx|^2 = \left|e^{\frac{-iE_n t_1}{\hbar}} \int_{-\infty}^{\infty} \psi_m^*(x)\psi_n(x)dx\right|^2 = \delta_{mn}\begin{cases}1, & \text{if } m = n\\ 0, & \text{if } m \neq n\end{cases}$

  Thus, the second measurement of energy will still yield $E_n$, and the result is independent of $t_1$. This means that the measurement of energy immediately after or a long time after the first energy measurement will yield the same value.
- **If we measure position at $t = t_1$**, the probability density for measuring $x_0$ is

  $\rho(x_0) = |\Psi(x_0,t_1)|^2 = \left|e^{\frac{-iE_n t_1}{\hbar}}\psi_n(x_0)\right|^2 = |\psi_n(x_0)|^2$, which is independent of $t_1$. This means that the measurement of position immediately after or a long time after the energy measurement will have the same probability density for measuring $x_0$.

➢ **If we measure position at $t = 0$ and obtain $x_0$**, the wavefunction will instantaneously collapse to $\delta(x - x_0)$, which can be expanded to $\sum_n C_n \psi_n(x)$ (in which $C_n = \psi_n^*(x_0)$). After time $t_1$, the wave function becomes $\sum_n C_n e^{\frac{-iE_n t_1}{\hbar}}\psi_n(x)$.

$$\overset{t=0}{\Psi(x,0) = \delta(x - x_0) = \sum_n C_n \psi_n(x)} \text{ - - - - - - - -}\rightarrow \overset{t=t_1}{\Psi(x,t_1) = \sum_n C_n e^{\frac{-iE_n t_1}{\hbar}}\psi_n(x)}$$

- **If we measure position at $t = t_1$,** the probability density for measuring $x_0$ is

  $\rho(x_0) = |\Psi(x_0,t_1)|^2 = \left|\sum_n C_n e^{\frac{-iE_n t_1}{\hbar}}\psi_n(x_0)\right|^2$, which depends on $t_1$. This means that the measurement of position immediately after or a long time after the first position measurement will generally have different probability density for measuring $x_0$.
- **If we measure energy at $t = t_1$**, the probability of obtaining $E_m$ is

  $P(E_m) = |\int_{-\infty}^{\infty} \psi_m^*(x)\Psi(x,t_1)dx|^2 = \left|\sum_n C_n e^{\frac{-iE_n t_1}{\hbar}} \int_{-\infty}^{\infty} \psi_m^*(x)\psi_n(x)dx\right|^2 =$

  $\left|\sum_n C_n e^{\frac{-iE_n t_1}{\hbar}}\delta_{mn}\right|^2 = |C_m|^2 = |\psi_m(x_0)|^2$, which is independent of $t_1$. This means that the measurement of energy immediately after or a long time after the position measurement will have the same probability of obtaining $E_m$.

**Note:**

It is important for instructors to discuss with students subtleties with regard to measurements of position. The eigenstates of the position operator are Dirac delta functions, which are not strictly normalizable or physical in the sense that position is not infinitely resolved in any true measurement. A Dirac delta function can be expressed as an infinite series of energy eigenstates, each of which has an infinitesimally small amplitude. Students should understand the unrealistic aspects of working with delta functions, even though they have pedagogical value and are convenient approximations for spatially localized states.

## Overall Class Discussion/Summary

If the particle is in an energy eigenstate $\psi_n(x)$ at $t = 0$

- Measure energy at $t = 0$:
  Probability of obtaining $E_n$ is 1 (and the probability would be the same if the measurement was made at $t = t_1$ instead of at $t = 0$)

- Measure position at $t = 0$:
  Probability density for measuring $x_0$ is $|\psi_n(x_0)|^2$ (and the probability density would be the same if the measurement was made at $t = t_1$ instead of at $t = 0$)

If at $t = 0$, the particle is in a state $\Psi(x)$ which is not an eigenstate of either $\hat{H}$ or any operator that commutes with $\hat{H}$.

- Measure energy at $t = 0$:
  Probability of obtaining $E_n$ is $|\int_{-\infty}^{\infty} \psi_n^*(x)\Psi(x)dx|^2$ (and the probability would be the same if the measurement was made at $t = t_1$ instead of at $t = 0$).

- Measure position at $t = 0$:
  Probability density for measuring $x_0$ is $|\Psi(x_0)|^2$ (and the probability density would be different if the measurement was made at $t = t_1$ instead of at $t = 0$). The position operator does not commute with the Hamiltonian, so position is not a constant of motion. The probability density for measuring position in a non-energy eigenstate depends on time.

## Appendix B Questions in the Pre-Test and Post-Test (Version A)

The pre- and post-test questions are reproduced below. The same information provided at the beginning of Appendix A applies to the questions in Appendix B.

*All of the questions in this test refer to an isolated system in which a particle is in a 1-D infinite square well with Hamiltonian* $\hat{H} = \frac{\hat{p}^2}{2m} + V(x)$ $(V(x) = 0\ when\ 0 < x < a, V(x) = +\infty\ otherwise)$. *The energy eigenvalues are* $E_n = \frac{n^2\pi^2\hbar^2}{2ma^2}$ $(n = 1,2,3,\ldots)$, *and the energy eigenstate corresponding to* $E_n$ *is* $\Psi_n(x) = \sqrt{\frac{2}{a}} sin\left(\frac{n\pi x}{a}\right)$ *when 0<x<a and* $\Psi_n(x) = 0$ *elsewhere.*

Q1. Choose all of the following statements that are correct about a generic state $\Psi(x)$ (which is not a stationary state).

I. $\hat{H}\Psi(x) = E_n\psi_n(x)$

II. $\hat{x}\Psi(x) = x\Psi(x)$

III. $\Psi(x)$ can be expressed as a linear superposition of the energy eigenstates.

Q2. The state of a particle at t=0 is given by $\Psi(x,0) = \sqrt{\frac{2}{7}}\Psi_1(x) + \sqrt{\frac{5}{7}}\Psi_2(x)$.

(a) If you measure the energy of the particle at t = 0, what possible energies could you obtain and what is the probability of each? Explain.

(b) If you measure the position of the particle at t = 0, what possible values could you obtain, and what is the corresponding probability density? Explain.

Q3. The state of a particle at t=0 is given by $\Psi(x,0) = \sqrt{\frac{2}{7}}\Psi_1(x) + \sqrt{\frac{5}{7}}\Psi_2(x)$

(a) If you measure energy at t = 0 and obtain a value of $E_1$, what is the normalized state of the system right after the measurement?

(b) Immediately after the measurement of energy in 3(a), you measure energy again. What is the probability of obtaining $E_1$?

(c) A long time after the measurement of energy in 3(a), you measure energy again. Will the probability of obtaining $E_1$ be the same or different as in 3(b)? Explain your reasoning.

(d) Immediately after the measurement of energy in 3(a), you measure position. What is the probability density of finding the particle at x = $x_0$? Explain.

(e) A long time after the measurement of energy in 3(a), you measure position. Will the probability density of finding the particle at x =$x_0$ be the same or different as in 3(d)? Explain your reasoning.

Q4. The state of a particle at t=0 is given by $\Psi(x,0) = \sqrt{\frac{2}{7}}\Psi_1(x) + \sqrt{\frac{5}{7}}\Psi_2(x)$

(a) If you measure position and obtain a value of $x_0$, what is the wavefunction of the system right after the measurement?

(b) Immediately after the measurement of position in 4(a), you measure position again. Will the wavefunction of the system right after the measurement be the same or different as in 4(a)? Explain.
(c) A long time after the measurement of position in 4(a), you measure position again. Will the wavefunction of the system right after the measurement be the same or different as in 4(b)? Explain your reasoning.
(d) Immediately after the measurement of position in 4(a), you measure energy. What is the probability of obtaining $E_1$? Explain.
(e) A long time after the measurement of position in 4(a), you measure energy. Will the probability of obtaining $E_1$ be the same or different as in 4(d)? Explain your reasoning.

[1] P. Jolly, D. Zollman, S. Rebello, and A. Dimitrova, Visualizing potential energy diagrams, Am. J. Phys 66, 57 (1998).

[2] A. Kohnle, M. Douglass, T. J. Edwards, A. D. Gillies, C. A. Hooley, and B. D. Sinclair, Developing and evaluating animations for teaching quantum mechanics concepts, Eur. J. Phys. 31, 1441 (2010).

[3] C. Manogue, E. Gire, D. McIntyre, and J. Tate, Representations for a spins-first approach to quantum mechanics, AIP Conf. Proc. 1413, 55 (2012).

[4] M. Wawro, K. Watson, and W. Christensen, Students' metarepresentational competence with matrix notation and Dirac notation in quantum mechanics, Phys. Rev. Phys. Educ. Res. 16, 020112 (2020).

[5] C. Singh, Student understanding of quantum mechanics, Am. J. Phys. 69, 885 (2001).

[6] D. Domert, C. Linder, and Å. Ingerman, Probability as a conceptual hurdle to understanding one-dimensional quantum scattering and tunnelling, Eur. J. Phys. 26, 47 (2004).

[7] G. Passante, P. J. Emigh, and P. S. Shaffer, Examining student ideas about energy measurements on quantum states across undergraduate and graduate levels, Phys. Rev. ST Phys. Educ. Res. 11, 020111 (2015).

[8] P. J. Emigh, G. Passante, and P. S. Shaffer, Student understanding of time dependence in quantum mechanics, Phys. Rev. ST Phys. Educ. Res. 11, 020112 (2015).

[9] C. Singh, Student understanding of quantum mechanics at the beginning of graduate instruction, Am. J. Phys. 76, 277 (2008).

[10] T. Tu, C.-F. Li, Z.-Q. Zhou, and G.-C. Guo, Students' difficulties with partial differential equations in quantum mechanics, Phys. Rev. Phys. Educ. Res. 16, 020163 (2020).

[11] P. Bitzenbauer, Effect of an introductory quantum physics course using experiments with heralded photons on preuniversity students' conceptions about quantum physics, Phys. Rev. Phys. Educ. Res. 17, 020103 (2021).

[12] R. Müller and H. Wiesner, Teaching quantum mechanics on an introductory level, Am. J. Phys. 70, 200 (2002).

[13] A. Kohnle, I. Bozhinova, D. Browne, M. Everitt, A. Fomins, P. Kok, G. Kulaitis, M. Prokopas, D. Raine, and E. Swinbank, A new introductory quantum mechanics curriculum, Eur. J. Phys. 35, 015001 (2013).

[14] E. Gire and E. Price, Structural features of algebraic quantum notations, Phys. Rev. ST Phys. Educ. Res. 11, 020109 (2015).

[15] K. Krijtenburg-Lewerissa, H. J. Pol, A. Brinkman, and W. Van, Joolingen, Insights into teaching quantum mechanics in secondary and lower undergraduate education, Phys. Rev. Phys. Educ. Res. 13, 010109 (2017).

[16] U. S. di Uccio, A. Colantonio, S. Galano, I. Marzoli, F. Trani, and I. Testa, Design and validation of a two-tier questionnaire on basic aspects in quantum mechanics, Phys. Rev. Phys. Educ. Res. 15, 010137 (2019).

[17] M. Michelini and A. Stefanel, in The International Handbook of Physics Education Research: Learning Physics, edited by M. F. Taşar and P. R. L. Heron (AIP Publishing LLC Online, Melville, NY, 2023).

[18] L. Branchetti, A. Cattabriga, and O. Levrini, Interplay between mathematics and physics to catch the nature of a scientific breakthrough: The case of the blackbody, Phys. Rev. Phys. Educ. Res. 15, 020130 (2019).

[19] G. Zhu and C. Singh, Improving student understanding of addition of angular momentum in quantum mechanics, Phys. Rev. ST Phys. Educ. Res. 9, 010101 (2013).

[20] B. Brown, A. Mason, and C. Singh, Improving performance in quantum mechanics with explicit incentives to correct mistakes, Phys. Rev. Phys. Educ. Res. 12, 010121 (2016).
[21] C. Singh, Interactive learning tutorials on quantum mechanics, Am. J. Phys. 76, 400 (2008).
[22] E. Marshman and C. Singh, Interactive tutorial to improve student understanding of single photon experiments involving a Mach–Zehnder interferometer, Eur. J. Phys. 37, 024001 (2016).
[23] C. Keebaugh, E. Marshman, and C. Singh, Improving student understanding of a system of identical particles with a fixed total energy, Am. J. Phys. 87, 583 (2019).
[24] S. DeVore and C. Singh, Interactive learning tutorial on quantum key distribution, Phys. Rev. Phys. Educ. Res. 16, 010126 (2020).
[25] S. Pollock, G. Passante, and H. Sadaghiani, Adaptable research-based materials for teaching quantum mechanics, Am. J. Phys. 91, 40 (2023).
[26] E. Marshman and C. Singh, Improving student understanding of Dirac notation by using analogical reasoning in the context of a three-dimensional vector space, presented at PER Conf. 2020, virtual conference, 10.1119/ perc.2020.pr.Marshman.
[27] C. Singh and E. Marshman, Investigating student difficulties with Dirac notation, presented at PER Conf. 2013, Portland, OR, 10.1119/perc.2013.pr.074.
[28] E. Marshman and C. Singh, Student difficulties with quantum states while translating state vectors in Dirac notation to wave functions in position and momentum representations, presented at PER Conf. 2015, College Park, MD, 10.1119/perc.2015.pr.048.
[29] C. Singh, Student difficulties with quantum mechanics formalism, AIP Conf. Proc. 883, 185 (2007).
[30] S.-Y. Lin and C. Singh, Categorization of quantum mechanics problems by professors and students, Eur. J. Phys. 31, 57 (2010).
[31] C. Singh and E. Marshman, Review of student difficulties in upper-level quantum mechanics, Phys. Rev. ST Phys. Educ. Res. 11, 020117 (2015).
[32] E. Marshman and C. Singh, Framework for understanding the patterns of student difficulties in quantum mechanics, Phys. Rev. ST Phys. Educ. Res. 11, 020119 (2015).
[33] C. Singh, M. Belloni, and W. Christian, Improving students' understanding of quantum mechanics, Phys. Today 8, 59, 43 (2006).
[34] E. Marshman and C. Singh, Investigating and improving student understanding of quantum mechanical observables and their corresponding operators in Dirac notation, Eur. J. Phys. 39, 015707 (2018).
[35] E. Marshman and C. Singh, Investigating and improving student understanding of the expectation values of observables in quantum mechanics, Eur. J. Phys. 38, 045701 (2017).
[36] E. Marshman and C. Singh, Investigating and improving student understanding of the probability distributions for measuring physical observables in quantum mechanics, Eur. J. Phys. 38, 025705 (2017).
[37] E. Marshman and C. Singh, Investigating and improving student understanding of quantum mechanics in the context of single photon interference, Phys. Rev. Phys. Educ. Res. 13, 010117 (2017).
[38] R. Sayer, A. Maries, and C. Singh, Advanced students' and faculty members' reasoning about the double slit experiment with single particles, presented at PER Conf. 2020, virtual conference, 10.1119/perc.2020.pr.Sayer.
[39] A. Maries, R. Sayer, and C. Singh, Can students apply the concept of "which-path" information learned in the context of Mach–Zehnder interferometer to the double-slit experiment?, Am. J. Phys. 88, 542 (2020).

[40] P. Hu, Y. Li, R. Mong, and C. Singh, Student understanding of the Bloch sphere, Eur. J. Phys. 45, 025705 (2024).
[41] R. Sayer, A. Maries, and C. Singh, Quantum interactive learning tutorial on the double-slit experiment to improve student understanding of quantum mechanics, Phys. Rev. Phys. Educ. Res. 13, 010123 (2017).
[42] C. Keebaugh, E. Marshman, and C. Singh, Investigating and addressing student difficulties with a good basis for finding perturbative corrections in the context of degenerate perturbation theory, Eur. J. Phys. 39, 055701 (2018).
[43] C. Keebaugh, E. Marshman, and C. Singh, Challenges in sense-making and reasoning in the context of degenerate perturbation theory in quantum mechanics, Phys. Rev. Phys. Educ. Res. 20, 020139 (2024).
[44] C. Keebaugh, E. Marshman, and C. Singh, Improving student understanding of fine structure corrections to the energy spectrum of the hydrogen atom, Am. J. Phys. 87, 594 (2019).
[45] C. Keebaugh, E. Marshman, and C. Singh, Improving student understanding of corrections to the energy spectrum of the hydrogen atom for the Zeeman effect, Phys. Rev. Phys. Educ. Res. 15, 010113 (2019).
[46] G. Zhu and C. Singh, Improving students' understanding of quantum measurement. I. Investigation of difficulties, Phys. Rev. ST Phys. Educ. Res. 8, 010117 (2012).
[47] G. Zhu and C. Singh, Improving students' understanding of quantum measurement. II. Development of researchbased learning tools, Phys. Rev. ST Phys. Educ. Res. 8, 010118 (2012).
[48] G. Zhu and C. Singh, Students' difficulties with quantum measurement, AIP Conf. Proc. 1413, 387 (2012).
[49] P. Hu, Y. Li, and C. Singh, Investigating and improving student understanding of the basics of quantum computing, Phys. Rev. Phys. Educ. Res. 20, 020108 (2004).
[50] P. J. Emigh, G. Passante, and P. S. Shaffer, Developing and assessing tutorials for quantum mechanics: Time dependence and measurements, Phys. Rev. Phys. Educ. Res. 14, 020128 (2018).
[51] D. J. Griffiths and D. F. Schroeter, Introduction to Quantum Mechanics (Cambridge University Press, 2018).
[52] P. Hu, Y. Li, and C. Singh, Challenges in addressing student difficulties with time-development of two-state quantum systems using a multiple-choice question sequence in virtual and in-person classes, Eur. J. Phys. 43, 025704 (2022).
[53] E. Marshman and C. Singh, Validation and administration of a conceptual survey on the formalism and postulates of quantum mechanics, Phys. Rev. Phys. Educ. Res. 15, 020128 (2019).
[54] G. Zhu and C. Singh, Surveying students' understanding of quantum mechanics in one spatial dimension, Am. J. Phys. 80, 252 (2012).
[55] P. R. L. Heron, Empirical investigations of learning and teaching, part I: Examining and interpreting student thinking, in Research on Physics Education: Proceedings of the International School of Physics "Enrico Fermi" Course CLVI (IOS Press, Amsterdam, 2004), Vol. 156, p. 341.
[56] E. Mazur, Peer Instruction: A User's Manual (Prentice Hall, Upper Saddle River, NJ, 1997).
[57] C. H. Crouch and E. Mazur, Peer Instruction: Ten years of experience and results, Am. J. Phys. 69, 970 (2001).
[58] C. Singh and G. Zhu, Improving students' understanding of quantum mechanics by using peer instruction tools, AIP Conf. Proc. 1413, 77 (2012).

[59] L. Ding, N. W. Reay, A. Lee, and L. Bao, Are we asking the right questions? Validating clicker question sequences by student interviews, Am. J. Phys. 77, 643 (2009).
[60] P. Justice, E. Marshman, and C. Singh, Development, validation and in-class evaluation of a sequence of clicker questions on Larmor precession of spin in quantum mechanics, presented at PER Conf. 2018, 10.1119/ perc.2019.pr.Justice.
[61] P. Justice, E. Marshman, and C. Singh, Improving student understanding of quantum mechanics underlying the SternGerlach experiment using a research-validated multiplechoice question sequence, Eur. J. Phys. 40, 055702 (2019).
[62] P. Justice, E. Marshman, and C. Singh, Student understanding of Fermi energy, the Fermi–Dirac distribution and total electronic energy of a free electron gas, Eur. J. Phys. 41, 015704 (2020).
[63] R. Sayer, E. Marshman, and C. Singh, Case study evaluating Just-In-Time Teaching and Peer Instruction using clickers in a quantum mechanics course, Phys. Rev. Phys. Educ. Res. 12, 020133 (2016).
[64] P. Hu, Y. Li, and C. Singh, Challenges in addressing student difficulties with measurement uncertainty of two-state quantum systems using a multiple-choice question sequence in online and in-person classes, Eur. J. Phys. 44, 015702 (2023).
[65] P. Hu, Y. Li, and C. Singh, Challenges in addressing student difficulties with basics and change of basis for two-state quantum systems using a multiple-choice question sequence in online and in-person classes, Eur. J. Phys. 44, 065703 (2023).
[66] P. Hu, Y. Li, and C. Singh, Challenges in addressing student difficulties with quantum measurement of two-state quantum systems using a multiple-choice question sequence in online and in-person classes, Phys. Rev. Phys. Educ. Res. 19, 020130 (2023).
[67] D. Schwartz, J. Bransford, and D. Sears, Efficiency and innovation in transfer, in Transfer of Learning from a Modern Multidisciplinary Perspective, edited by J. Mestre (Information Age Publishing, Greenwich, CT, 2005), Vol. 3, p. 1.
[68] D. L. Schwartz and J. D. Bransford, A time for telling, Cognit. Instr. 16, 475 (1998).
[69] T. J. Nokes-Malach and J. P. Mestre, Toward a model of transfer as sense-making, Educ. Psychol. 48, 184 (2013).
[70] K. Shabani, M. Khatib, and S. Ebadi, Vygotsky's zone of proximal development: Instructional implications and teachers' professional development, Eng. Lang. Teach. 3, 237 (2010).
[71] E. Hutchins, Enculturating the supersized mind, Philos. Stud. 152, 437 (2011).
[72] E. Hutchins, The cultural ecosystem of human cognition, Philos. Psychol. 27, 34 (2014).
[73] P. Heller, R. Keith, and S. Anderson, Teaching problem solving through cooperative grouping. Part 1: Group versus individual problem solving, Am. J. Phys. 60, 627 (1992).
[74] C. Singh, Impact of peer interaction on conceptual test performance, Am. J. Phys. 73, 446 (2005).
[75] M. J. Brundage, A. Malespina, and C. Singh, Peer interaction facilitates co-construction of knowledge in quantum mechanics, Phys. Rev. Phys. Educ. Res. 19, 020113 (2023).
[76] A Ghimire and C. Singh, How often does unguided peer interaction lead to correct response consensus? An example from Conceptual Survey of Electricity and Magnetism, Eur. J. Phys. 45, 035703 (2024).
[77] A. A. DiSessa, A friendly introduction to "knowledge in pieces": Modeling types of knowledge and their roles in learning, in Invited Lectures from the 13th International Congress on Mathematical Education, ICME-13 Monographs (Springer, Cham, 2018), p. 65.
[78] C. Keebaugh, E. Marshman, and C. Singh, Investigating and addressing student difficulties with the corrections to the energies of the hydrogen atom for the strong and weak field Zeeman effect, Eur. J. Phys. 39, 045701 (2018).

[79] C. Keebaugh, E. Marshman, and C. Singh, Investigating and improving student understanding of the basics for a system of non-interacting identical particles, Am. J. Phys. 90, 110 (2022).
[80] J. Cohen, Statistical Power Analysis for the Behavioral Sciences (L. Erlbaum Associates, Hillsdale, NJ, 1988).
[81] R. R. Hake, Interactive-engagement versus traditional methods: A six-thousand-student survey of mechanics test data for introductory physics courses, Am. J. Phys. 66, 64 (1998).
[82] J. M. Nissen, R. M. Talbot, A. Nasim Thompson, and B. Van Dusen, Comparison of normalized gain and Cohen's d for analyzing gains on concept inventories, Phys. Rev. Phys. Educ. Res. 14, 010115 (2018).
[83] E. Mazur, Peer instruction: Getting students to think in class, AIP Conf. Proc. 399, 981 (1997).
[84] E. Gire and C. Manogue, Making sense of quantum operators, eigenstates and quantum measurements, AIP Conf. Proc. 1413, 195 (2012).